\documentclass[useAMS,usenatbib]{mn2e}
\usepackage{amsmath,graphicx,bm,amsbsy,aas_macros,url,array,amssymb}
\usepackage[caption = false]{subfig}
\def\spose#1{\hbox to 0pt{#1\hss}}
\def\simlt{\mathrel{\spose{\lower 3pt\hbox{$\mathchar"218$}}
     \raise 2.0pt\hbox{$\mathchar"13C$}}}
\def\simgt{\mathrel{\spose{\lower 3pt\hbox{$\mathchar"218$}}
     \raise 2.0pt\hbox{$\mathchar"13E$}}}
\def\simpropto{\mathrel{\spose{\lower 3pt\hbox{$\mathchar"218$}}
     \raise 2.0pt\hbox{$\propto$}}}

\def\ie{{\frenchspacing\it i.e.}}

\def\spose#1{\hbox to 0pt{#1\hss}}
\def\simlt{\mathrel{\spose{\lower 3pt\hbox{$\mathchar"218$}}
   \raise 2.0pt\hbox{$\mathchar"13C$}}}
\def\simgt{\mathrel{\spose{\lower 3pt\hbox{$\mathchar"218$}}
     \raise 2.0pt\hbox{$\mathchar"13E$}}}
 \def\simpropto{\mathrel{\spose{\lower 3pt\hbox{$\mathchar"218$}}
     \raise 2.0pt\hbox{$\propto$}}}

\def\beq#1{\begin{equation}\label{#1}}
\def\eeq{\end{equation}}
\def\beqa#1{\begin{eqnarray}\label{#1}}
\def\eeqa{\end{eqnarray}}

\def\fig#1{Figure~\ref{#1}}
\def\Fig#1{Figure~\ref{#1}}

\title{MITEoR: A Scalable Interferometer for Precision 21~cm Cosmology}
\author[H.~Zheng, et al.]{H.~Zheng$^{1}$\thanks{E-mail:
jeff\_z@mit.edu}, M.~Tegmark$^{1}$, V.~Buza$^{1,2}$, J.~S.~Dillon$^{1}$, H.~Gharibyan$^{1}$, J.~Hickish$^{3}$,
\newauthor E.~Kunz$^{1}$, A.~Liu$^{1,4,5}$, J.~Losh$^{1}$, A.~Lutomirski$^{1}$, S.~Morrison$^{1}$, S.~Narayanan$^{1}$,
\newauthor A.~Perko$^{1,6}$, D.~Rosner$^{1}$, N.~Sanchez$^{1}$, K.~Schutz$^{1}$, S.~M.~Tribiano$^{1,7}$, M.~Valdez$^{1,8}$,
\newauthor H.~Yang$^{1}$, K.~Zarb Adami$^{3}$, I.~Zelko$^{1}$, K.~Zheng$^{1}$, R.~P.~Armstrong$^{1,12}$, R.~F.~Bradley$^{9,10}$, 
\newauthor M.~R.~Dexter$^{4}$, A.~Ewall-Wice$^{1}$, A.~Magro$^{11}$, M.~Matejek$^{1}$, E.~Morgan$^{1}$,
\newauthor A.~R.~Neben$^{1}$, Q.~Pan$^{1}$, R.~F.~Penna$^{1}$, C.~M.~Peterson$^{2}$, M.~Su$^{1}$, J.~Villasenor$^{1}$, 
\newauthor C.~L.~Williams$^{1}$, Y.~Zhu$^{1,6}$ \\
$^{1}$Dept. of Physics and MIT Kavli Institute, Massachusetts Institute of Technology, 77 Massachusetts Ave., Cambridge, MA 02139, USA\\
$^{2}$Dept.~of Physics, Harvard University, Cambridge, MA 02138, USA\\
$^{3}$Dept.~of Physics, University of Oxford, Oxford, OX1 3RH, United Kingdom\\
$^{4}$Dept.~of Astronomy and Radio Astronomy Lab, University of California, Berkeley, CA 94720, USA\\
$^{5}$Berkeley Center for Cosmological Physics, Berkeley, CA 94720, USA\\
$^{6}$Dept.~of Physics, Stanford University, Stanford, CA 94305, USA\\
$^{7}$Science Dept.~Borough of Manhattan Community College, City University of New York, New York, NY 10007, USA\\
$^{8}$Dept.~of Astronomy, Boston University, Boston, MA 02215, USA\\
$^{9}$Dept.~of Electrical and Computer Engineering, University of Virginia, Charlottesville, VA 22904, USA\\
$^{10}$National Radio Astronomy Observatory, Charlottesville, VA 22903, USA\\
$^{11}$Dept.~of Physics, University of Malta, Msida MSD 2080, Malta\\
$^{12}$Dept.~of Astronomy, University of Cape Town, Private Bag X3, Rondebosch 7701, South Africa}
\date{\today}
\begin{document}


\maketitle

\begin{abstract}
We report on the MIT Epoch of Reionization (MITEoR) experiment, a pathfinder low-frequency radio interferometer whose goal is to test technologies that improve the calibration precision and reduce the cost of the high-sensitivity 3D mapping required for 21\,cm cosmology. MITEoR accomplishes this by using massive baseline redundancy, which enables both automated precision calibration and correlator cost reduction. We demonstrate and quantify the power and robustness of redundancy for scalability and precision. We find that the calibration parameters precisely describe the effect of the instrument upon our measurements, allowing us to form a model that is consistent with $\chi^2$ per degree of freedom $< 1.2$ for as much as $80\%$ of the observations. We use these results to develop an optimal estimator of calibration parameters using Wiener filtering, and explore the question of how often and how finely in frequency visibilities must be reliably measured  to solve for calibration coefficients.
The success of MITEoR with its 64 dual-polarization elements bodes well for the more ambitious Hydrogen Epoch of Reionization Array (HERA) project and other next-generation instruments, which would incorporate many identical or similar technologies.
\end{abstract} 


\maketitle

\section{Introduction}

Mapping neutral hydrogen throughout our universe via its redshifted 21~cm line 
offers a unique opportunity to probe the cosmic ``dark ages," the formation of the first luminous objects, and the epoch of reionization (EoR). A suitably designed instrument with a tenth of a square kilometer of collecting area will allow tight constraints on the timing and duration of reionization and the astrophysical processes that drove it  \citep{poberliudillon}. 
Moreover, because it can map a much larger comoving volume of our universe, it has the potential to overtake the 
Cosmic Microwave Background (CMB) as our most sensitive cosmological probe of inflation, dark matter, dark energy, and neutrino masses. For example  \citep{Yi}, a radio array with a square kilometer of collecting area, maximal sky coverage, and good foreground maps could improve the sensitivity to spatial curvature and neutrino masses by up
to two orders of magnitude, to $\Delta\Omega_k\approx 0.0002$ and $\Delta m_\nu\approx 0.007$ eV, and shed new light on the early universe by a $4\sigma$ detection of the spectral index running predicted by the simplest inflation models favored by the BICEP2 experiment  \citep{bicep2}.

Unfortunately, the cosmological 21~cm signal is so faint that none of the current experiments around the world (LOFAR  \citealt{LOFAR}, MWA  \citealt{MWA}, PAPER  \citealt{PAPER}, 21CMA  \citealt{21cma}, GMRT  \citealt{GMRT})
have detected it yet, although increasingly stringent upper limits have recently been placed  \citep{GMRT2,MWAJosh,PAPERpspec}. A second challenge is that foreground contamination from our galaxy and extragalactic sources is perhaps four orders of magnitude larger than the cosmological hydrogen signal  \citep{GSM}. Any attempt to accurately clean it out from the data requires even greater sensitivity as well as more accurate calibration and beam modeling than the current state-of-the-art in radio astronomy (see  \citet{FurlanettoReview,miguelreview} for reviews).

Large sensitivity requires large collecting area. Since steerable single dish radio telescopes become prohibitively expensive beyond a certain size, the aforementioned experiments have all opted for interferometry, combining $N$ (generally a large number) independent antenna elements which are (except for GMRT) individually more affordable. The LOFAR, MWA, PAPER, 21CMA and GMRT experiments currently have comparable $N$. The problem with scaling interferometers to high $N$ is that  all of these experiments use standard hardware cross-correlators whose cost grows quadratically with $N$, since they need to correlate all $N(N-1)/2\sim N^2/2$ pairs of antenna elements. This cost is reasonable for the current scale $N\sim 10^2$, but will completely dominate the cost for $N\simgt 10^3$, making precision cosmology arrays with $N\sim 10^6$ as discussed in  \citet{Yi} infeasible in the near future, which has motivated novel correlator approaches such as  \citet{moff}.

For the particular application of 21~cm cosmology, however, designs with better cost scaling are possible, as described in  \citet{FFTT,FFTT2}: by arranging the antennas in a 
hierarchical rectangular or hexagonal grid and performing the correlations using Fast Fourier Transforms (FFTs), thereby cutting the cost scaling to $N\log N$. This is particularly attractive for science applications requiring exquisite sensitivity at vastly different angular scales, such as
21~cm cosmology (where short baselines are needed to probe the cosmological signal\footnote{It has been shown that the 21~cm signal-to-noise ratio ($S/N$) per resolution element in the $uv$-plane (Fourier plane) is $\ll 1$ for all current 21~cm cosmology experiments, and that their cosmological sensitivity therefore improves by moving their antennas closer together to focus on the center of the $uv$-plane and bringing its $S/N$ closer to unity
 \citep{Morales2005,Bowman2006,Matt3,Yi,Lidz2009}. Error bars on the cosmological power spectrum have contributions from both noise and sample variance, and it is well-known that the total error bars on a given physical scale (for a fixed experimental cost) are minimized when both contributions are comparable, which happens when the $S/N\sim 1$ on that scale. This is why more compact 21~cm experiments have been advocated.
This is also why early suborbital CMB experiments focused on small patches of sky to get $S/N\sim 1$ per pixel, and why galaxy redshift surveys target objects like luminous red galaxies that give $S/N\sim 1$ per 3D voxel.
} 
and long baselines are needed for point source removal).
Such hierarchical grids thus combine the angular resolution advantage of traditional array layouts with the cost advantage of a rectangular Fast Fourier Transform Telescope. If the antennas have a broad spectral response as well and their signals are digitized with high bandwidth, the cosmological neutral hydrogen gets simultaneously imaged in a vast 3D volume covering both much of the sky and also a vast range of distances (corresponding to different redshifts, \ie, different observed frequencies.)
Such low-cost arrays have been called {\it omniscopes} \citep{FFTT,FFTT2} for their wide field of view and broad spectral range.

Of course, producing such scientifically rich maps with any interferometer depends crucially on our ability to precisely calibrate the instrument, so that we can truly understand how our measurements relate to the sky. Traditional radio telescopes rely on a well-sampled Fourier plane to perform self-calibration using the positions and fluxes of a number of bright point sources. At first blush, one might think that any highly-redundant array would be at a disadvantage in its attempt to calibrate the gains and phases of individual antennas. However, we can use the fact that redundant baselines should measure the same Fourier component of the sky to improve the calibration of the array dramatically and quantifiably. In fact, we find that the ease and precision of redundant baseline calibration is a strong rationale for building a highly-redundant array, in addition to the improvements in sensitivity and correlator speed.

Redundant calibration is useful both for current generation redundant arrays like MITEoR and PAPER and for future large arrays that will need redundancy to cut down correlator cost. Omniscopes must be calibrated in real time, because they do not compute and store the visibilities measured by each pair of antennas,  but effectively gain their speed advantage by averaging redundant baselines in real time. Individual antennas therefore cannot be calibrated in post-processing. No calibration scheme used on existing low frequency radio interferometers has been demonstrated to meet the speed and precision requirements of omniscopes. Thus, the main goal of the MIT Epoch of Reionization experiment (MITEoR) and this paper is to demonstrate a successful redundant calibration pipeline that can overcome the calibration challenges faced by current and future generation instruments by performing automatic precision calibration in real time.

Building on past redundant baseline calibration methods by Wieringa  \citep{Wieringa} and others, some of us recently developed an algorithm which is both automatic and statistically unbiased, able to produce precision phase and gain calibration for all antennas in a hierarchical grid (up to a handful of degeneracies) without making any assumptions about the sky signal  \citep{omnical}. Once obtained, precision calibration solutions can in turn produce more accurate modeling of the synthesized and primary beams\footnote{For tile-based interferometers like the MWA and 21CMA, gain and phase errors in individual antennas (as opposed to tiles) do not typically get calibrated in the field, adding a fundamental uncertainty to the tile sky response.}  \citep{JonniePrimaryBeam}, which has been shown to improve the quality of the foreground modeling and removal which is so crucial to 21~cm cosmology. It is therefore timely to develop a pathfinder instrument that tests how well the latest calibration ideas works in practice. 

MITEoR is such a pathfinder instrument, designed to test redundant baseline calibration. We developed and successfully applied a real-time redundant calibration pipeline to data we took with our 64 dual-polarization antenna array during the summer of 2013 in The Forks, Maine. The goal of this paper is to describe the design of the MITEoR instrument, demonstrate the effectiveness of our redundant baseline calibration and absolute calibration pipelines, and use the calibration results to obtain an optimal scheme for estimating calibration parameters as a function of time and frequency. 

This paper is organized as follows. We first describe in Section \ref{secinstrument} the instrument, including the custom developed analog components, the 8 bit 128 antenna-polarization correlator, the deployment, and the observation history. 
In Section \ref{seccal}, we focus on precision calibration.
We explain and quantitatively evaluate relative redundant calibration, and address the question of how often calibration coefficients should be updated. We also examine the absolute calibration, including breaking the degeneracies in relative calibration, mapping the primary beam, and measuring the array orientation.
In Section \ref{secsummary}, we summarize this work and discuss implications for future redundant arrays such as HERA  \citep{poberliudillon}.
\section{The MITEoR Experiment}\label{secinstrument}
\begin{figure}
\centerline{\includegraphics[width=85mm]{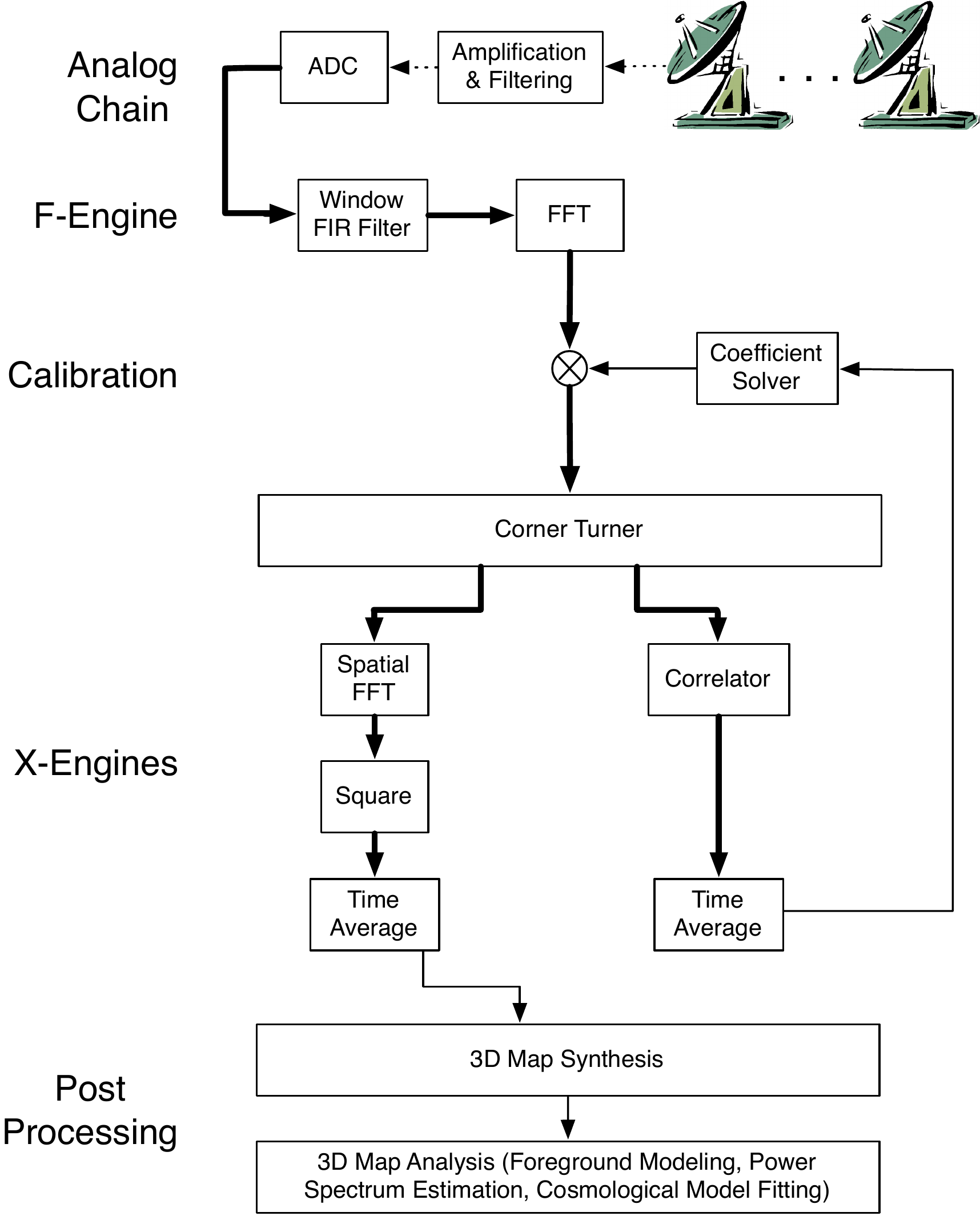}}
\caption{
Data pipeline for a large omniscope that implements FFT correlator and redundant baseline calibration. First, a hierarchical grid of dual-polarization antennas converts the sky signal into volts, which get amplified and filtered by the analog chain, transported to a central location, and digitized every few nanoseconds.
These high-volume digital signals (thick lines) get processed by field-programmable gate arrays (FPGAs) which perform a temporal Fourier transform. The FPGAs (or GPUs) then multiply by complex-valued calibration coefficients that depend on antenna, polarization and frequency, then spatially Fourier transform, square and accumulate the results, recording integrated sky snapshots every few seconds and thus reducing the data rate by a factor $\sim 10^9$.  They also cross-correlate a small fraction of all antenna pairs, allowing the redundant baseline calibration software  \citep{omnical,LOFARcal} to update the calibration coefficients in real time and automatically monitor the quality of calibration solutions for instrumental malfunctions.
Finally, software running on regular computers combine all snapshots of sufficient quality into a 3D sky ball or ``data cube" representing the sky brightness as a function of angle and frequency in Stokes (I,Q,U,V)  \citep{FFTT2}, and subsequent software accounts for foregrounds and measures power spectra and other cosmological observables.
\label{ArchitectureFig}
}
\end{figure}

In theory, a very large omniscope can be built following the generalized architecture in \fig{ArchitectureFig}. On the other hand, it is crucial to demonstrate that automatic and precise calibration is possible in real-time using redundant baselines, since the calibration coefficients for each antenna must be updated frequently to allow the FFTs to combine the signals from the different antennas without introducing errors. In this section, we will present our partial implementation of this general design, including both the analog and the digital systems. Because the digital hardware is powerful enough to allow it, the MITEoR prototype correlates all 128 input channels with one another, rather than just a small sample as mentioned in the caption of \fig{ArchitectureFig}. This provides additional cross-checks that greatly aid technological development, where instrumentation may be particularly prone to systematics. This also allows us to explore the question of exactly how often and how finely in frequency we must measure visibilities to solve for calibration coefficients, a question we return to in Section \ref{seccal}. Since we chose to implement a full correlator, an additional FFT correlator would bring no extra information (simply computing the same redundant-baseline-averaged visibilities faster), so we leave the digital implementation of an FFT correlator to future work.  In general, our mission is to empirically explore any challenges that are unique to a massively redundant interferometer array.  Once these are known, one can reconfigure the cross-correlation hardware to perform spatial FFTs, thereby obtaining an omniscope with $N\log N$ correlator scaling.

\subsection{The Analog System}\label{secanalog}

\begin{figure}
\centerline{\includegraphics[width=85mm]{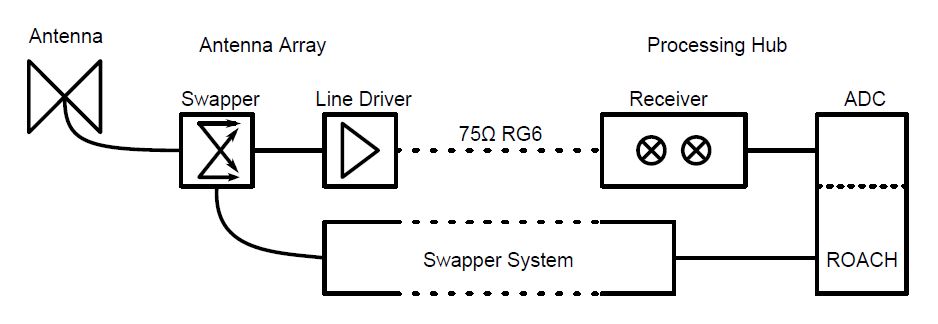}}
\caption{
System diagram of the analog system. The signal received with an MWA ``bow-tie" antenna is first amplified by the built-in low noise amplifier, then Walsh-modulated in the swapper module controlled by the swapper system. The signal is amplified again in the line driver and sent to the processing rack through 50\,m long coaxial cables. In the processing rack, the signal first goes into the receiver, where it undergoes further amplification, frequency down-mixing and I/Q modulation from the 120-180\,MHz range to the 0-25\,MHz range. The analog chain ends with digitization on ADC connected to ROACH boards. \label{analog_schematic}
}
\end{figure}

MITEoR contains 64 dual-polarization antennas, giving 128 signal channels in total. The signal picked up by the antennas is first amplified by two orders of magnitude in power by the low noise amplifiers (LNAs) built-in to the antennas. It is then phase switched in the swapper system, which greatly reduces cross-talk downstream. The signal is then amplified again by about five orders of magnitude in the line-drivers before being sent over 50 meter RG6 cables to the receivers. The receivers perform IQ demodulation on a desired $50\,\textrm{MHz}$ band selected between $100\,\textrm{MHz}$ and $200\,\textrm{MHz}$, producing two channels with adjacent $25\,\textrm{MHz}$  bands, and sends the resulting signals into the digitization boards containing 256 analog-to-digital converters (ADCs) sampling at $50\,\textrm{MHz}$.  The swappers, line-drivers and receivers we designed are shown in \fig{analog_schematic}.

When designing the components of this system, we chose to use commercially-available integrated circuits and filters whenever possible, to allow us to focus on
system design and construction. In some cases (such as with the amplifiers) the cost of the IC is less than the cost of enough discrete transistors to implement even a rough approximation of the same functionality. Less expensive filters
could be made from discrete components, but the characteristics of purchased modules tend to be  better due to custom inductors and shielding. When we needed to produce our own boards as described below, our approach was to design, populate and test them in our laboratory, then have them affordably mass-produced for us by Burns Industries\footnote{ \url{http://www.burnsindustriesinc.com}}.


\subsubsection{Antennas}
\begin{figure}
\centerline{\includegraphics[width=85mm]{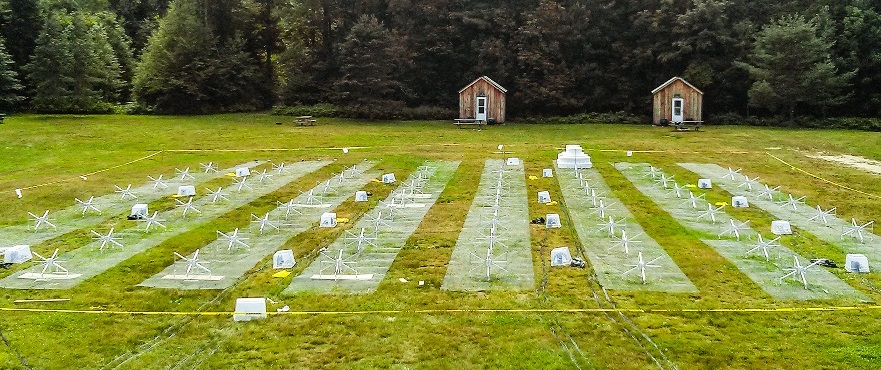}}
\caption{Part of the MITEoR array during the most recent deployment in the summer of 2013. 64 dual-polarization antennas were laid on a 21~m by 21~m regular grid with 3~m separation. The digital system was housed in the back of a shielded U-Haul truck (not shown). 
\label{ArrayFigure}
}
\end{figure}

The dual-polarization antennas used in MITEoR were originally developed for the Murchison Widefield Array  \citep{MWAantenna,MWA}, and consist of two ``bow-tie"-shaped arms as can be seen in \fig{ArrayFigure}. They are inexpensive, easy to assemble, and sensitive to the entire band of our interest. The MWA antennas were designed for the frequency range $80$-$300\,\textrm{MHz}$, and
have a built-in low noise amplifier with $20\,\textrm{dB}$ of gain. The noise figure of the amplifier is $0.2\,\textrm{dB}$, and the $20\,\textrm{dB}$
of gain means that subsequent gain stages do not contribute significantly to the noise figure\footnote{In a multi-stage amplifier, the contribution of each stage's noise figure is suppressed by a factor that is equal to the total gain of previous stages.}.

\subsubsection{Swappers (Phase Switches)}
\begin{figure*}
\centerline{\includegraphics[width=180mm]{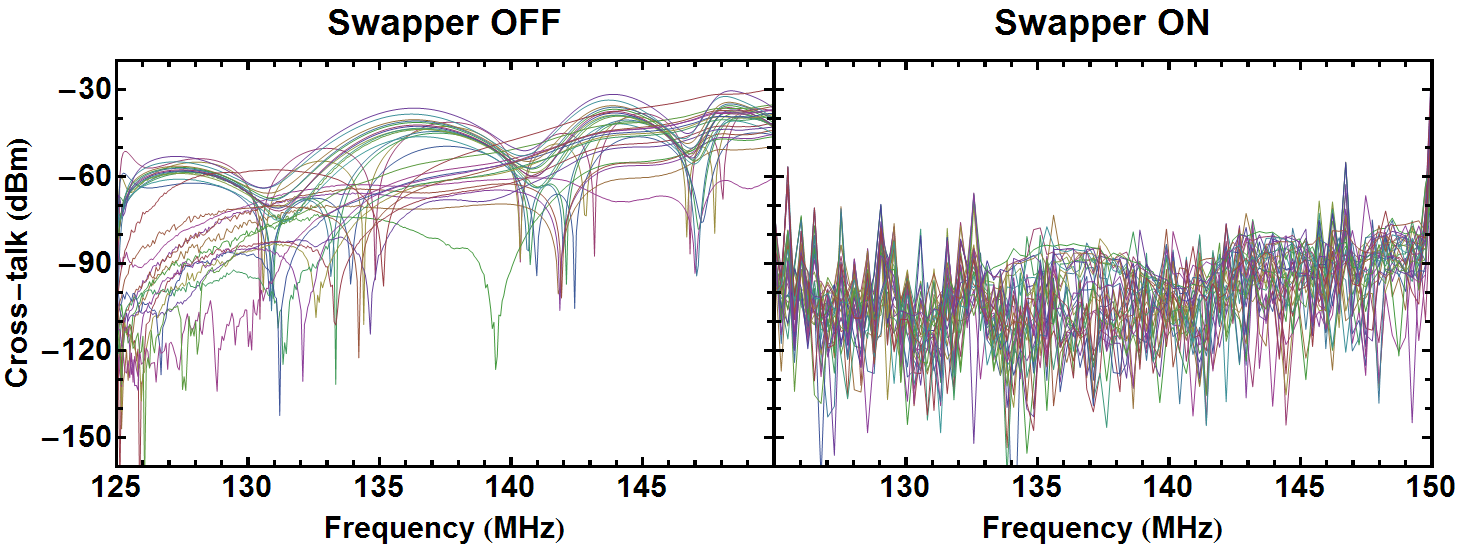}}
\vskip-2mm
\caption{
Plots of cross-talk power measured in the laboratory. The swapper suppresses crosstalk between channels by as much as 50\,dB. To measure these curves we fed a 0\,dBm sinusoidal signal into input channel 0 of the receivers and left the other 31 input channels open. We then measured the correlations between channel 0 and all 31 empty channels, due to crosstalk from channel 0. We repeated the procedure with input frequencies from 125-150\,MHz and obtained the results shown above.  
\label{SwapperFig}
}
\end{figure*}

As with many other interferometers, crosstalk within the receivers, ADCs, and cabling significantly affects signal quality. 
We observe the cross-talk to depend strongly on the physical proximity of channel pairs, reaching as high as about $-30$\,dB between nearest neighbor receiver channels. Our swapper system is designed to cancel out crosstalk during the correlator's time averaging by selectively inverting analog signals using Walsh modulation  \citep{nevadathesis}. The signal from each antenna-polarization is inverted 50\% of the time according to its own Walsh function, by an analog ZMAS-1 phase switch from Mini-Circuits located before the second amplification stage (line-driver), then appropriately re-inverted after digitization\footnote{Since the undesirable crosstalk signal is demodulated with a different Walsh function than it is modulated with, it will be averaged out due to orthogonality of Walsh functions.}. We perform the inversion once every millisecond, which is much longer than the ADC's 20\,ns sample time, and much shorter than the averaging time of a few seconds\footnote{The inversion cannot be too frequent, because we need to discard data during the analog inversion process which takes a few microseconds. At the same time, the inversion needs to be frequent enough to average out the cross-talk.}. This eliminates all crosstalk to first order  \citep{nevadathesis}.
\begin{figure}
\centerline{\includegraphics[width=80mm]{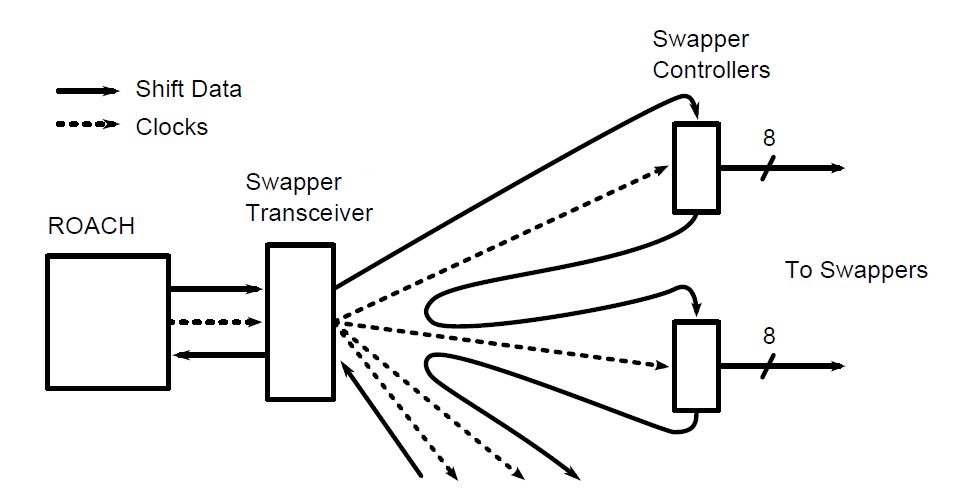}}
\vskip0.3cm
\centerline{\includegraphics[width=80mm]{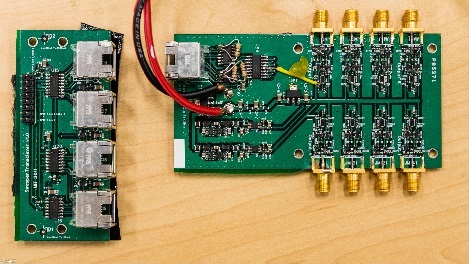}}
\caption{
System diagram of our swapper signal system and physical components of the swapper transceiver (lower left) and swapper controller (lower right). The swapper is designed to reduce crosstalk between neighboring channels.
\label{Swapper}
}
\end{figure}
If crosstalk reduction were the only concern, the ideal position for the swapper would be 
immediately after the antenna, in order to cancel as much crosstalk as possible. In practice, the swapper introduces a loss of about 3\,dB, so we perform the modulation after the LNA to avoid adding noise (raising the system temperature). To evaluate the effectiveness of the swapper modules, we sent a monotone signal into one single channel of the receivers while leaving other channels open, and measured the correlation between the signal channel and each empty channels with the swapper turned on and off. We then repeated this while varying the signal frequency over the full range of interest. As seen in \fig{SwapperFig},  
the swapper system attenuates crosstalk in the receiver and ADC by as much as 50\,dB over the frequency band of interest, typically reducing it to being of order $-80$\,dB for strongly afflicted signal pairs.

\subsubsection{Line-driver}

\begin{figure}
\vskip-2cm
\centerline{\includegraphics[width=110mm]{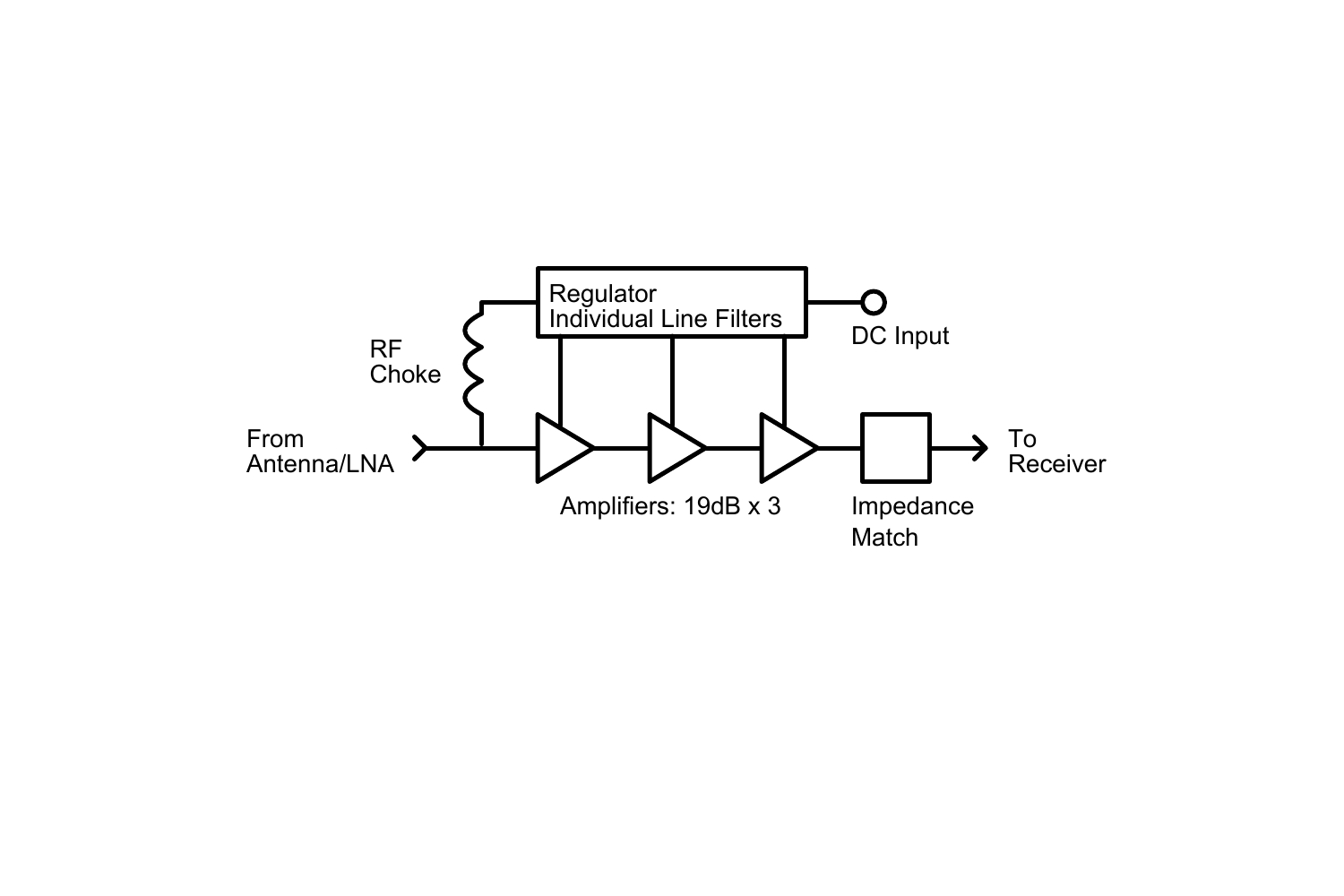}}
\vskip-2.7cm
\centerline{\includegraphics[width=80mm]{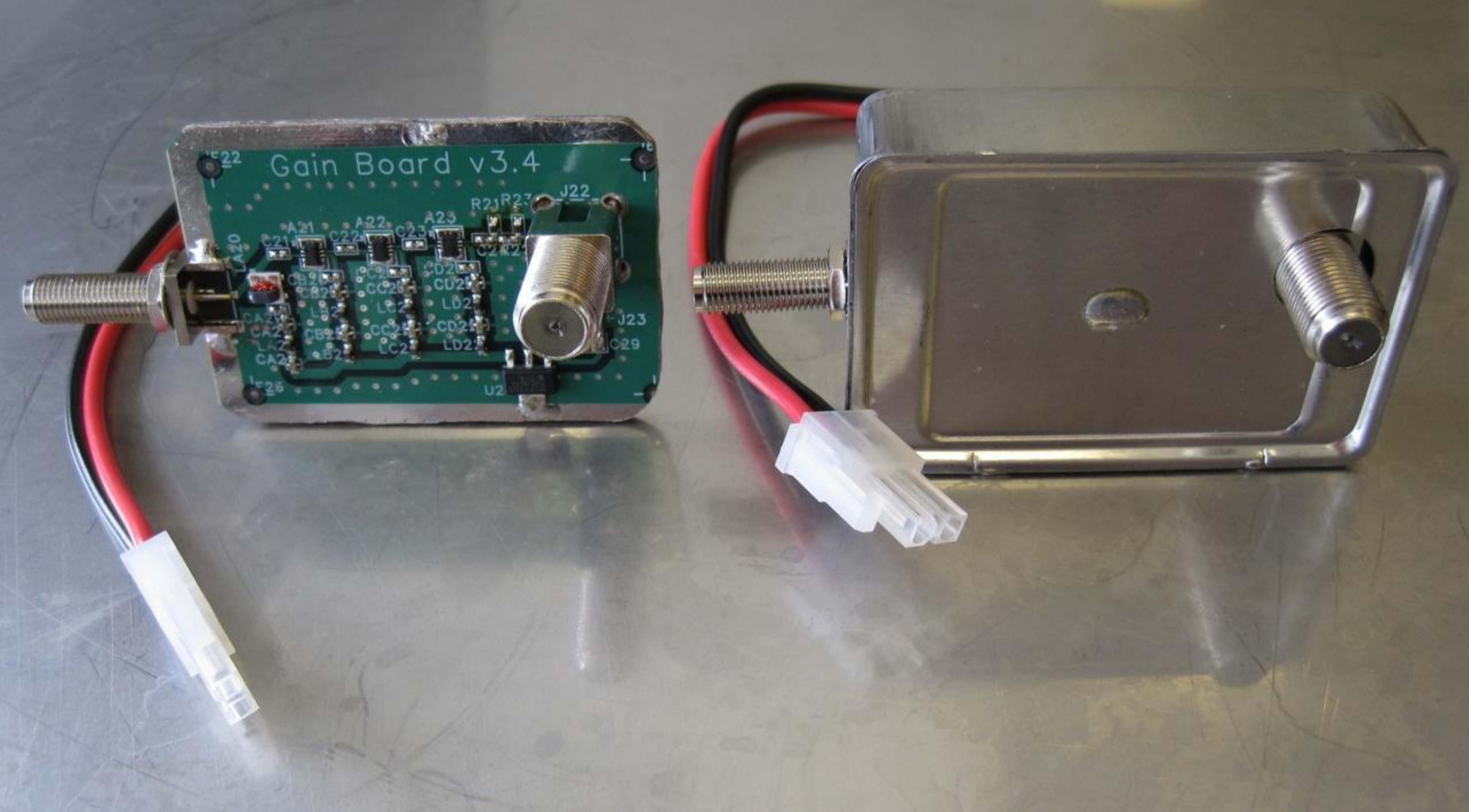}}
\caption{
System diagram and physical components of the line drivers. The line driver we designed takes the signal in the 50$\Omega$ coaxial cable from the antenna LNA and amplifies it by 51\,dB, in order to overpower noise picked up in the subsequent 75$\Omega$ coaxial cable and further processing steps up to 50 meters away.  It operates on 5V DC and also provides DC bias power to the antenna's LNA through the 50 Ohm cable.
\label{LineDriverFig}
}
\end{figure}

A line-driver (\fig{LineDriverFig}) amplifies a single antenna's signal from one of its two polarization channels while also powering its LNA. Line-drivers only handle a single channel to reduce potential crosstalk from sharing a printed circuit board. They are placed within a few meters of the antennas in order to reduce resistive
losses from powering the antenna at low voltage. Additional gain that they provide early in the analog chain helps the signal overpower
any noise picked up along the way to the processing hub, and maintains the low noise figure set up by the LNA. To further reduce potential radio-frequency interference (RFI), we chose to power the line-drivers with 58Ah 6V sealed lead acid rechargeable batteries during the final 64-antenna deployment, rather than 120\,VAC to 6\,VDC adapters (whose unwanted RF-emission may have caused occasional saturation problems during our earlier expeditions).

\subsubsection{Receiver}
\begin{figure}
\vskip-1.3cm
\centerline{\includegraphics[width=90mm]{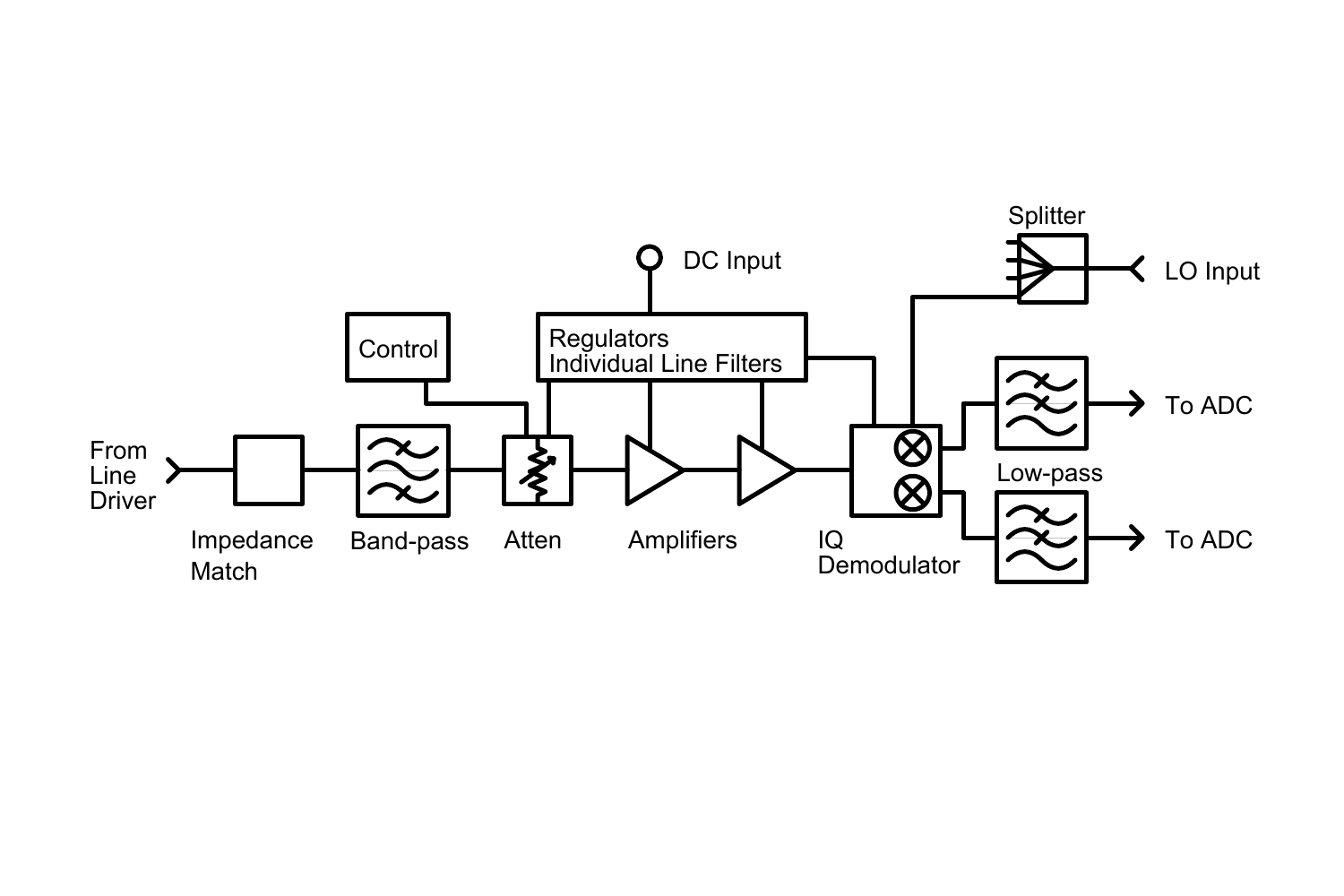}}
\vskip-1.9cm
\centerline{\includegraphics[width=80mm]{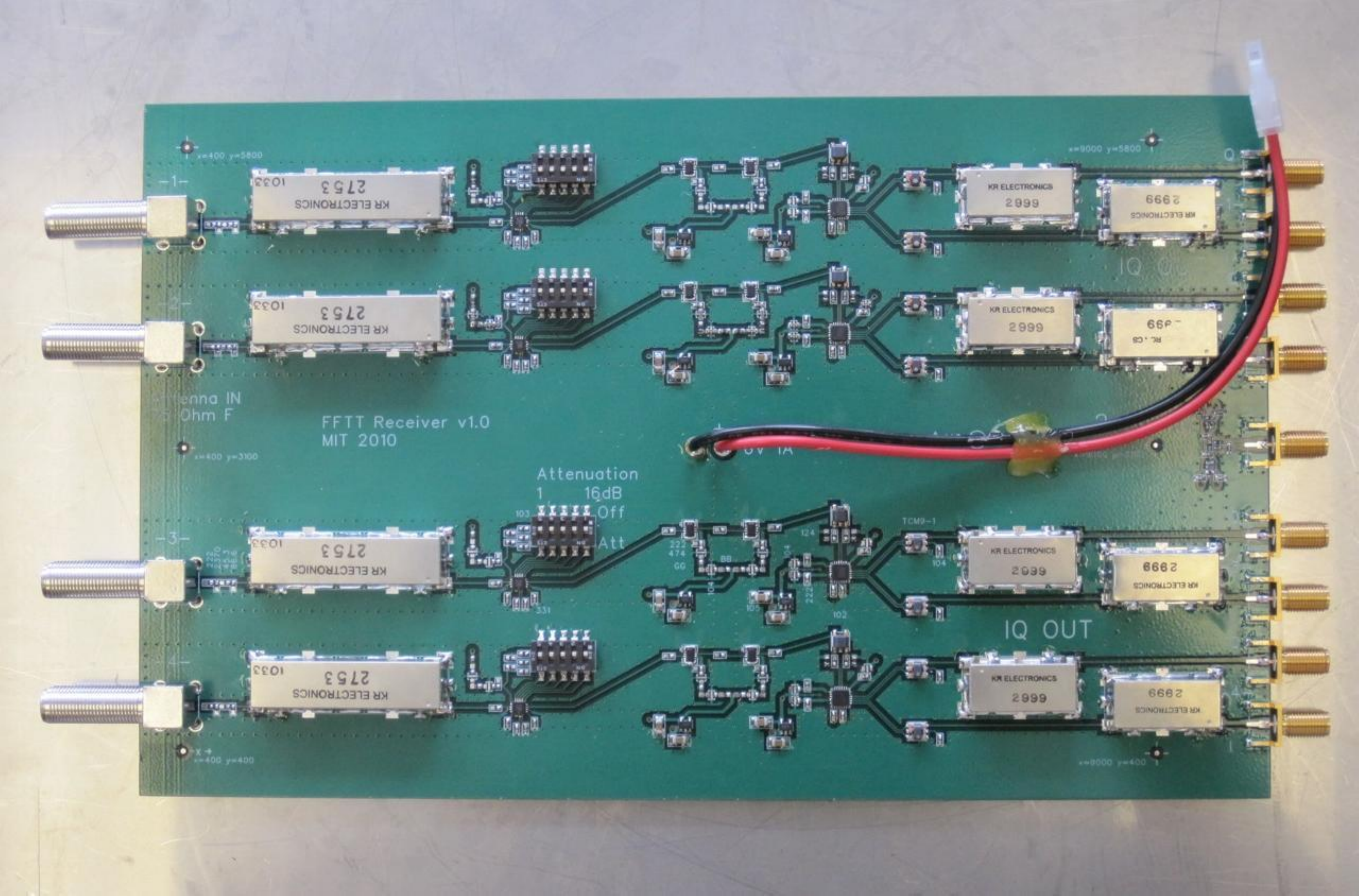}}
\caption{
System diagram and physical component of the receiver boards. The boards take the signals arriving from four line drivers, 
band-pass filter and amplify them, then use a local oscillator to frequency shift them from the band of interest to a DC-centered signal suitable for input to the ADC.
\label{ReceiverFig}
}
\end{figure}

Our receivers (\fig{ReceiverFig}) take input from the line-drivers, bandpass filter the incoming signals, amplify their power level by 23\,dB, and IQ-demodulate them. The resulting signals go directly to an ADC for digitization. Receivers are placed near the ADCs to which they are connected to reduce cabling for local oscillator (LO) distribution and ADC connections. 
IQ demodulation is used, which doubles received bandwidth for a given ADC frequency at the cost of using two ADC channels, and has the advantage of requiring only a single  LO and low speed ADCs. The result is 40 MHz of usable bandwidth\footnote{Due to limitations in our FPGAs' computing power, only 12.5\,MHz of digitized data are correlated and stored at any instant.} anywhere in the range 110-190 MHz, with a 2-3 MHz gap centered around the LO frequency due to bandpass filters.
The receiver boards have five pins allowing their signals to be attenuated by any amount between 0\,dB and 31\,dB (in steps of 1\,dB) before the second amplification stage, to avoid saturation and non-linearities from RFI and to attain signal levels optimal for digitization.

\subsection{The Digital System}\label{secdigital}
\label{digi}
\begin{figure}
\centerline{\includegraphics[width=80mm]{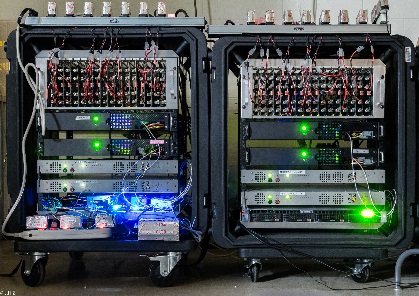}}
\caption{
The entirety of our 128 antenna-polarization digital correlator system, packaged in two portable shock mounted racks. The two black chassis and two silver chassis in the middle of each rack are F-engines (ROACH) and X-engines (ROACH2), respectively. Above the ROACHes are 32 receiver boards that input the signals from 128 line drivers via F-cables. The blue lit area below the ROACHes contains various clocking devices responsible for synchronization, whereas the chassis below the ROACHes on the right hand side is the 8TB data acquisition server.
\label{RackFig}
}
\end{figure}

We designed MITEoR's digital system (\fig{RackFig}) to be highly compact and portable. The entire system occupies 2 shock-mounted equipment racks on wheels, each measuring about 1\,m on all sides. It takes in data from 256 ADC channels (64 antennas with I and Q signals for polarizations), Fourier transforms each channel, reconstructs IQ demodulated channels back to 128 corresponding antenna channels, computes the cross-correlations of all pairs of the 128 antenna channels with 8 bit precision, and then time-averages these cross-correlations. Although standard 4 bit correlators suffice for most astronomical observation tasks, the better dynamic range of our 8 bit correlator allows us to observe faint astronomical signals at the same time as $10^3$ times brighter ORBCOMM satellites, whose enormous signal-to-noise has proved invaluable in characterizing various aspects of the system (see Sections \ref{secbeam}, \ref{secrotation}, and \ref{secsystematic}). The digital hardware is capable of processing an instantaneous bandwidth of $12.5\,\textrm{MHz}$ with $49\,\textrm{kHz}$ frequency bins. It averages those correlations and then writes them to disk every few seconds (usually either 2.6 or 5.3 seconds).

While one of the advantages of a massively redundant interferometer array is the ability to reduce costs by performing a spatial FFT rather than a full cross-correlation, we have not implemented FFT correlation in the current MITEoR prototype as the hardware is powerful enough to correlate all antenna pairs in real time (the feasibility of implementing FFT correlation on the ROACH platform has been demonstrated by  \citet{best2fftt}). Rather, the goal of MITEoR is to quantify the accuracy that automatic redundant baseline calibration can attain, thereby experimentally characterizing all of the unknowns in the system, such as unexpected analog chain systematics and other barriers to finding good calibration solutions.

We adopted the widely-used F-X scheme in MITEoR's digital system. We have 4 synchronized F-engines that take in data from 4 synchronized 64-channel ADC boards, which run at 12 bits and 50Ms/s. The F-engines perform the FFT and IQ reconstruction, and distribute the data onto 4 X-engines through 16 10GbE links. The 4 asynchronous X-engines each perform full correlation on 4 different frequency bands on all 128 antenna polarizations, and send the time averaged results to a computer for data storage.

To implement the computational steps of the MITEoR design, we used Field Programmable Gate Arrays (FPGAs). These devices can be programmed to function as dedicated pieces of computational hardware. Each F-engine and X-engine is implemented by one Xilinx FPGA (Virtex-5 for F-engines and Virtex-6 for X-engines). These FPGAs are seated on custom hardware boards developed by the CASPER collaboration\footnote{\url{https://casper.berkeley.edu/}}  \citep{4176933}. We also use the software tool flow developed by CASPER to design the digital system. The CASPER collaboration is dedicated to building open-source programmable hardware specifically for applications in astronomy. We currently use two of their newer devices, the ROACH\footnote{\url{https://casper.berkeley.edu/wiki/ROACH}} (Reconfigurable Open Architecture Computing Hardware) for the F-engines, and the ROACH 2\footnote{\url{https://casper.berkeley.edu/wiki/ROACH-2_Revision_2}} for the X-engines. The main benefit of using CASPER hardware is that it facilitates the time-consuming process of designing
and building custom radio interferometry hardware. The CASPER collaboration also offers a large open-source library of FPGA blocks for commonly used signal processing structures such as polyphase filter banks, FIR filters and fast Fourier transform blocks  \citep{4840623}. However, due to MITEoR's ambitious architecture, involving both extreme compactness, an 8-bit correlator, and tight inter-ROACH synchronization constraints, we custom-designed most of the digital FPGA blocks. The specifications of our latest correlator are listed in Table \ref{tabspecs}.

\subsection{MITEoR Deployment and Data Collection}\label{secdeploy}
\begin{table}
\begin{tabular}{ | m{3.7cm} | m{3.9cm} | }
	\hline
	\vskip0.5mm
	Antenna & MWA dual-pol bow-tie \\ \hline
	\vskip0.5mm
	Antenna count & 64 $\times$ 2 polarizations \\ \hline
	\vskip0.5mm
	Array configuration &  8 $\times$ 8 grid  \\ \hline
	\vskip0.5mm
	ADC & 4 $\times$ 64-channel 50\,Msps \\ \hline
	\vskip0.5mm
	F-engine & 4 ROACHes with Virtex-5\\ \hline
	\vskip0.5mm
	X-engine & 4 ROACH2s with Virtex-6\\ \hline
	\vskip0.5mm
	Correlator precision &  8 bits  \\ \hline
	\vskip0.5mm
	Frequency range & 110-190\,MHz \\ \hline
	\vskip0.5mm
	Instantaneous bandwidth & 12.5\,MHz  (50 MHz digitized)\\ \hline
	\vskip0.5mm
	Frequency resolution & 49\,kHz \\ \hline
	\vskip0.5mm
	Time resolution & $\geq$ 2.68\,s \\ \hline
  \end{tabular}
\caption{
List of MITEoR specifications. We observed with two different 8 by 8  array configurations, one with 3\,m separation and one with 1.5\,m separation. We observed ORBCOMM band with 2.68\,s resolution, and we chose a resolution of 5.37\,s for other bands. 
\label{tabspecs}
}
\end{table}
We deployed MITEoR in The Forks, Maine, which our online research suggested might be one of the most radio quiet region in the United States at the frequencies of interest\footnote{The Forks has also been successfully used to test the EDGES experiment  \citep{EDGES}, and we found the RFI spectrum to be significantly cleaner than at the National Radio Astronomy Observatory in Green Bank, West Virginia at the very low (100-200 MHz) frequency range that is our focus: the entire spectrum at The Forks is below -100\,dBm except for one -89.5\,dBm spike at 150MHz.}\citep{rfi}. We deployed the first prototype in September 2010, and performed a successful suite of test observations with an 8-antenna interferometer. In May 2012, we completed and deployed a major upgrade of the digital system to fully correlate $N=16$ dual-polarization antennas. With the experience of this successful deployment, we further upgraded the digital system to accommodate $N=64$ dual-polarization antennas, which led to our latest deployment in July 2013 and the results we describe in this paper.

The MITEoR experiment was designed to be portable and easy to assemble. The entire experiment was loaded into a 17 foot U-Haul truck and driven to The Forks. It took a crew of 15 people less than 2 days to assemble the instrument and bring it to full capacity. A skeleton crew of 3 members stayed on site for monitoring and maintenance for the following two weeks, during which we collected more than 300 hours of data. Subsequently, a demolition crew of 5 members disassembled and packed up MITEoR in 6 hours and concluded the successful deployment.

During the deployment, we scanned through the frequency range $123.5$-$179.5\,\textrm{MHz}$, with at least 24 consecutive hours at each frequency. We used two different array layouts for most of the frequencies we covered. The observation began with the antennas arranged in a regular 8 by 8 grid, with 3 meter spacing\footnote{We aligned the antenna positions using a laser-ranging total station, and measured their positions with millimeter level precision. The median deviation from a perfect grid is 2\,mm in the N-S direction, 3\,mm E-W, and 28\,mm vertical, primarily caused by the fact that the deployment site had not been leveled.} between neighboring antennas, which we later reconfigured to an 8 by 8 regular grid with 1.5 meter spacing for a more compact layout (which provides better signal-to-noise ratio on the 21~cm signal). The total volume of binary data collected was 3.9TB, and in the rest of this paper, we demonstrate the results of our various calibration techniques using this data set.

\section{Calibration Results}\label{seccal}

As we have emphasized above, the precision calibration of an interferometer is essential to its ability to detect the faint cosmological imprint upon the 21~cm signal, and the key focus of MITEoR is to determine how well real-time redundant calibration can be made to work in practice. In this section we describe the calibration scheme that we have designed and implemented and quantify its performance. We first constrain the {\it relative} calibration between antennas, utilizing both per-baseline algorithms and redundant-baseline calibration algorithms  \citep{omnical}. We then build on these relative calibration results to constrain the {\it absolute} calibration of the instrument, including breaking the few degeneracies inherent to redundant calibration.

\subsection{Relative Calibration}\label{secrelcal}

\subsubsection{Overview}
The goal of relative calibration is to calibrate out differences among antenna elements caused by non-identical analog components, such as variations in amplifier gains and cable lengths, which may be functions of time and frequency. We parametrize the calibration solution as a time- and frequency-dependent multiplicative complex gain $g_i$ for each of the 128 antenna-polarizations. Calibrating the interferometer amounts to solving for the coefficients $g_i$ and undoing their effects on the data.
\begin{figure*}
\hskip-5mm\centerline{\includegraphics[width=190mm]{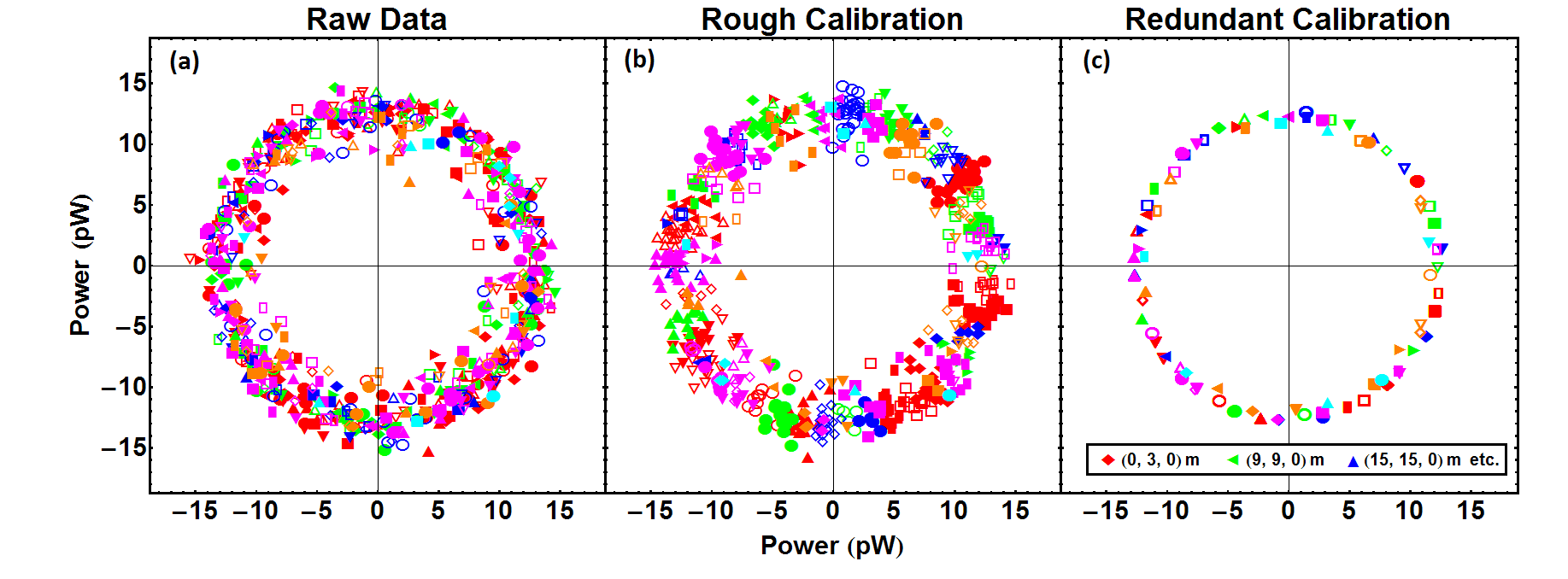}}
\caption{
Illustration of three stages in the redundant baseline calibration pipeline. Each panel is a complex plane, and each point is a complex visibility for a specific pair of antennas at 137.1\,MHz during the passage of an ORBCOMM satellite. Each unique combination of color and shape stands for one set of redundant baselines. In an ideal world, all identical symbols, such as all upright red triangles, should have the same value thus overlap exactly. Due to noise, they should cluster together around the same complex value. In panel (a) showing raw data, the redundant baselines have almost no clustering visible---for example, red filled circles can be found throughout the plot. After crude calibration in panel (b), we see most points falling into clustered segments---though the clustering is still far from exact. Finally in panel (c), after performing log calibration, we see that all points corresponding to each redundant baseline are almost exactly overlapping, with no visible deviation due to the high signal-to-noise. While the difference is not visible here, linear calibration can further improve log calibration results, as shown in \fig{chisq}.
\label{omniviewer}
}
\end{figure*}
Our calibration scheme revolves around calibration methods that heavily utilize the redundancy of our array, whose efficacy we aim to demonstrate with MITEoR. The current redundant calibration pipeline involves three steps, as illustrated in \fig{omniviewer}: 
\begin{enumerate}
\item Rough calibration computes approximate calibration phases using knowledge of the sky.
\item Logarithmic calibration (``logcal") decomposes roughly calibrated data into amplitudes and phases and computes least square fits for amplitude and phase separately.
\item Linear calibration (``lincal") takes the relatively precise but biased results from logcal and computes unbiased calibration parameters with even higher precision. 
\end{enumerate}
Although logcal and lincal have been previously proposed  \citep{Wieringa, omnical, lincal}, they both fail  in their original form if the phases of $g_i$ are not close to 0.\footnote{Logcal requires phase calibrations close to 0 to avoid phase wrapping issues, whereas lincal requires phase calibrations close to 0 to converge.} In practice, the phases of $g_i$ can be anywhere in the interval $[0, 2\pi)$. To overcome these practical challenges, we introduced various improvements to these algorithms. In the following sections, we describe our improvements to calibration algorithms in detail, and demonstrate the effectiveness of our calibration by obtaining $\chi^2/\mbox{DoF}\approx 1$ for the majority of our data. We then analyze these calibration parameters to construct a Wiener filter to optimally average them over time and frequency, which also tells us how frequently we need to calibrate in time and frequency.

\subsubsection{Rough Calibration}\label{secrough}
The goal of rough calibration is to obtain reliable initial phase estimates for the calibration parameters to enable the subsequent more sophisticated algorithms. This step does not have to involve redundancy, thus it can be done with any standard calibration techniques, for example self-calibration \citep{selfcalreview,selfcal}. The rough calibration algorithm that we describe below is computationally cheap and can robustly improve upon raw data even when a few antennas have failed.

At a given time and frequency, we have both the measured visibilities, $v_{ij}$, and $v^\text{model}_{ij}$, a rough model of the true sky signal\footnote{Since we are trying to obtain an initial estimate, the model does not have to be very accurate.}, where indices $i$ and $j$ represent antenna number. We first compute the phase difference between each measured visibility and its prediction. We then pick one reference antenna and subtract the phases of its visibilities from the phases of other visibilities to obtain a list of estimated phase calibration for each antenna. Finally, we take the median of these calibration phases to obtain a robust phase calibration estimator for each antenna. More concretely, we use the following procedure:
\begin{enumerate}
\item Construct a matrix $\mathbf{M}$ of phase differences where $M_{ij}=-M_{ji}= \arg(v_{ij}/v^\text{model}_{ij})$.
\item Define the first antenna as the reference by subtracting the first column of $\mathbf{M}$ from all columns to obtain $M'_{jk} = M_{jk} - M_{j0}$.
\item Obtain rough phase calibration parameters $\phi_k \equiv \arg(g_k)$ by computing the median angle of column $k$ in $M'$, defined as 
\begin{align}\label{eqmangle}
\phi_k
\equiv&\arg\,[\mbox{median}_j\{\exp(i M'_{jk})\}] \nonumber \\
=&\arg\,[\mbox{median}_j \{\cos(M'_{jk})\} \nonumber \\
&+i\,\mbox{median}_j \{\sin(M'_{jk})\}].
\end{align} 
\end{enumerate}
For stable instruments, the true calibration parameters have very small variation over days, so we can use one set of rough calibration parameters from a single snapshot in time for data from all other times. Thus we pick a snapshot at noon when each $v^\text{model}_{ij}$ can be easily computed from position of the Sun, and use the resulting raw calibration parameters as the starting point for logcal at all other times.

\subsubsection{Log Calibration and Linear Calibration}\label{secloglin}
To explain our redundant calibration procedure, we first need to briefly reintroduce the formalism developed in  \citet{omnical}. Suppose the $i^{\rm th}$ antenna measures a signal $s_i$ at a given instant. This signal can be expressed in terms of a complex gain factor $g_i$, the antenna's instrumental noise contribution $n_i$, and the true sky signal $x_i$ that would be measured in the limit of perfect gain and no noise:
\begin{equation}
s_i = g_i x_i + n_i.
\end{equation}
Under the standard assumption that the noise is uncorrelated with the signal, each baseline's measured visibility is the correlation between the two signals from the two antennas: 
\begin{align}
\label{eq:eq1}
v_{ij} &\equiv \langle s_i^* s_j \rangle \nonumber \\
&=g_i^* g_j\langle x_i^* x_j\rangle + g_i^*\langle x_i^* n_j\rangle + g_j\langle n_i^* x_j\rangle +\langle n_i^* n_j\rangle\nonumber \\ 
&= g_i^* g_jy_{i-j} + n_{ij}^{\text{res}},
\end{align}
where we have denoted the true correlation $\langle x_i^* x_j\rangle$ by $y_{i-j}$,\footnote{Following  \citet{omnical}, we use $y_{i-j}$ instead of $y_{ij}$ to emphasize that in a redundant array, the number of unique baseline visibilities can be much smaller than number of measured visibilities. The complete expression should be $y_{u(i,j)}$, where $u(i,j)$ means that baseline $ij$ corresponds to the $u$th unique baseline.} the noise from each antenna by $n_i$, the noise for each baseline by $n_{ij}^{\mbox{res}}$, and expectation values (effectively time averages) by angled brackets $\langle \dots \rangle$. In a maximally redundant array such as MITEoR, the number of unique baselines is much smaller than the total number of baselines. Therefore, we can treat all the $g_i$s and the $y_{i-j}$s as unknowns while keeping the system of equations \eqref{eq:eq1} overdetermined, enabling fits for both despite the presence of instrumental noise.

In  \citet{omnical}, some of us proposed logcal and lincal, and we have implemented both for calibrating MITEoR data. In log calibration, we take the logarithm of both sides of Equation \ref{eq:eq1} and obtain a linearized equation in logarithmic space. We then perform a least squares fit for the system of equations 
\begin{equation}
\log v_{ij} =  \log g_i^*+ \log g_j + \log y_{i-j},
\end{equation}
 where we solve for $\log g_i$ and $\log y_{i-j}$. Because the least squares fit takes place in log space whereas the noise is additive in linear space, the best fit results are biased. Linear calibration, on the other hand, is unbiased  \citep{omnical}. The lincal method performs a Taylor expansion of Equation \ref{eq:eq1} around initial estimates $g_{i}^0$ and $y_{i-j}^0$ and obtains a system of linearized equations
\begin{equation}
v_{ij} =  g_{i}^{0*} g_{j}^0y_{i-j}^0 +  g_i^{1*} g_j^0y_{i-j}^0+  g_i^{0*} g_j^1y_{i-j}^0+  g_i^{0*} g_j^0y_{i-j}^1,
\end{equation}
where we solve for $g_i^1$ and $y_{i-j}^1$.  For a detailed description of the logcal and lincal algorithms and their noise properties, we direct the reader to  \citet{omnical,lincal}. We now describe some essential improvements to these algorithms. 

Logcal was first thought to be unable to calibrate the phase component due to phase wrapping, since logcal has no way to recognize that $0^\circ$ and $360^\circ$ are the same quantity. Consider, for example a pair of redundant baselines that measure phases of $0.1^\circ$ and $359.9^\circ$ respectively. We can infer that they each only need a very small phase correction ($\pm0.1^\circ$) to agree perfectly. However, since logcal treats the difference between them as $359.8^\circ$ rather than $0.2^\circ$, it will calibrate by averaging $0.1^\circ$ and $359.9^\circ$ to $180^\circ$, which is completely wrong.

We made two improvements to the logcal method to guard against this. The first is to perform rough calibration beforehand, as described in Section \ref{secrough}. The second is to re-wrap the phases of $v_{ij}$. While rough calibration can make the phase errors relatively small\footnote{In our experience, they need to be less than about 20 degrees to ensure that the subsequent calibration steps converge reliably.}, that improvement alone is not sufficient, since $0^\circ$ and $360^\circ$ are still treated as different quantities. Thus we need to intelligently wrap the phases of the input vector before feeding it into logcal. This is done in two simple steps. For a snapshot of rough calibrated visibilities at given time and frequency, $v_{ij}$,  we first estimate the true phases of each group of redundant baselines, $\arg(y_{i-j})$, by computing median angles of measured phases using Eq. \ref{eqmangle}. Then for each measured phase, we add or subtract $2\pi$ until it is within $\pm\pi$ of $\arg(y_{i-j})$. This eliminates the phase wrapping problem. 

Unlike logcal, lincal is an unbiased algorithm, but it relies on a set of initial estimates for the correct calibration solutions to start with. The output of lincal can be fed back into the algorithm and it can iteratively improve upon its own solution.
However, the algorithm converges to the right answer only if the initial estimates are good. In practice, we find that three iterations of lincal typically produces excellent convergence, because the outputs of logcal are already decent estimates of the calibration solutions. Thus, by improving logcal, we also greatly improve lincal's effectiveness.

Our current calibration pipeline performs all steps of redundant calibration in less than 1 millisecond on a single processor core for a data slice at one time and one frequency channel, which is an order of magnitude faster than the rate data is saved onto disk. It is carried out by our open source \texttt{Omnical} package, coded in C++/Python.\footnote{The package supports the miriad file format and is easily adapted to work with other file formats. To obtain a copy, please contact jeff\_z@mit.edu.}
Thus there should be no computational challenge in performing the above described calibration procedure in real-time for any array with less than $10^3$ elements. For a future omniscope that has as many as $10^6$ elements, there are two ways to reduce the computational cost. The first is to calibrate less frequently in time and frequency, and we will discuss in detail the minimal sampling frequency in Section \ref{secwiener}. The other is to adapt a hierarchical redundant calibration scheme, where instead of calibrating all visibilities at the same time, one can calibrate the array in a hierarchical fashion whose computational cost scales only linearly with the number of elements. We discuss more details regarding hierarchical redundant calibration in Appendix \ref{apphierarchical}.
\begin{figure*}
\includegraphics[width = 180mm]{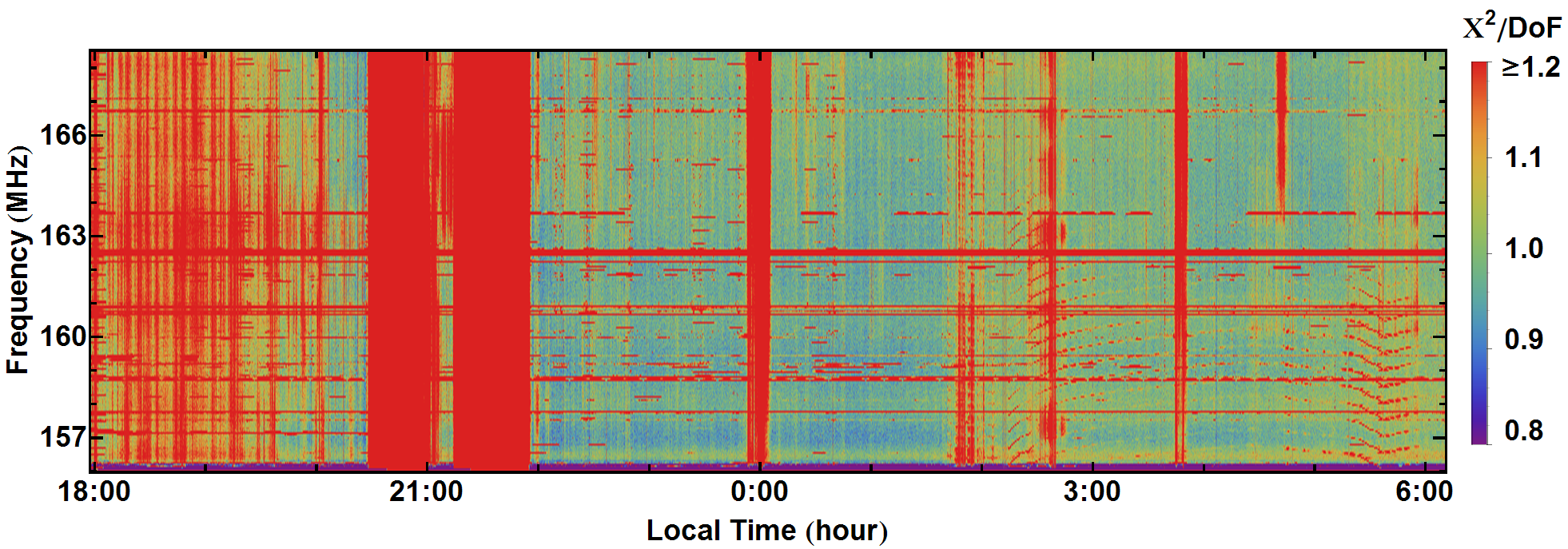}
\caption{
Waterfall plot of $\chi^2$/DoF for a day's data. This demonstrates the stability of our instrument as well as the effectiveness of using $\chi^2$/DoF as a indicator of data quality. We evaluate $\chi^2$/DoF every 5.3 seconds and every 49\,kHz. For the majority of the night time data, $\chi^2$/DoF is close to 1. We flag all data with $\chi^2$ larger than 1.2, which are marked red in this plot and account for 20\% of this data set. The amount of detailed structure in the flagged area (around 18:00 for example) shows the $\chi^2$ flagging technique's sensitivity to rapidly changing data quality.
\label{chisqwaterfall}
}
\end{figure*}

\subsubsection{$\chi^2$ and Quality of Calibration}\label{secchisq}

One of the many advantages of redundant calibration is it allows for the calculation of a $\chi^2$ for every snapshot to quantify how accurate the estimated visibilities are for each unique baseline, even without any knowledge of the sky. For a set of visibilities at a given time and frequency, $v_{ij}$, with calibration results $g_i$ and $y_{i-j}$, we define $\chi^2$ as
\begin{align}
\chi^2 & = \sum_{ij}{\frac{|v_{ij}-y_{i-j}g_i^*g_j|^2}{\sigma_{ij}^2}}\label{eqchisq},
\end{align}
where $\sigma_{ij}^2$ is the noise contribution to the variance of  the visibility $v_{ij}$. 
The effective number of degrees of freedom (DoF) is
\begin{align}
\mbox{DoF} & = N_{\text{measurements}} - N_{\text{parameters}}\nonumber\\
&= N_{\text{baselines}} - (N_{\text{antennas}} + N_{\text{unique baselines}}).
\end{align}
The numerator in Equation \ref{eqchisq} represents the deviation of measured data, $v_{ij}$, from the best fit redundant model, $y_{i-j}g_i^*g_j$. Thus $\chi^2$/DoF can be interpreted as the non-redundancy in measured data divided by the expected non-redundancy from pure noise. If the data agrees perfectly with the redundant model (with noise) and is free from systematics, then $\chi^2/\text{DoF}$ is drawn from a $\chi^2$ distribution with mean 1 and variance $2/\text{DoF}$ \citep{abramowitz1964handbook}.

\begin{figure}
\includegraphics[width = 83mm]{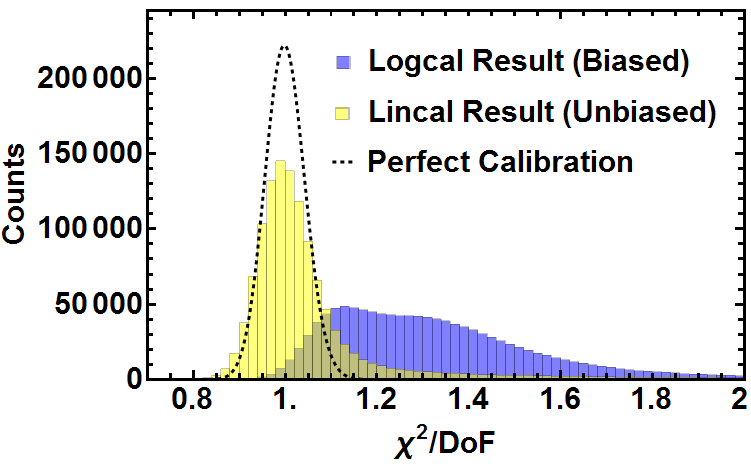}
\caption{
Histograms of the distributions of $\chi^2$/DoF of logcal results (mean 1.31) and lincal results (mean 1.05, median 1.01), together with the theoretical distribution of $\chi^2$/DoF (mean 1). They contain one night of data in a 12.5MHz frequency band (21:00-5:00 in \fig{chisqwaterfall}). We evaluate $\chi^2$/DoF for every 5.3 seconds and every 49\,kHz. We set the flagging threshold to $\chi^2=1.2$, and 80\% of the lincal result is below the threshold (majority of the 20\% flagged data have $\chi^2$ much larger than 2, thus not shown in this figure). Among the data that is not flagged, 85\% is accounted for by the theoretical $\chi^2$ distribution. The right tail in lincal's distribution is due to the noise model sometimes underestimating the noise in order to minimize false negatives in the flagging process. The fact that $\chi^2$/DoF for lincal is so close to the theoretical distribution means that both the instrument and the calibration algorithms are working exactly as we expect.
\label{chisq}
}
\end{figure}

With a smooth model for $\sigma_{ij}$ which we describe below, we compute $\chi^2$/DoF for the results of rough calibration, log calibration, and linear calibration using all of our data. The $\chi^2$ distributions of our calibrations for one day's data are shown in Figures~\ref{chisqwaterfall} and~\ref{chisq}. Each calibration algorithm significantly reduces the $\chi^2$/DoF, and lincal's produces a distribution of $\chi^2$/DoF consistently centered around 1. We automatically flag any data with $\chi^2$/DoF larger than 1.2, which accounts for about 20\% of the data. Among the data that is not flagged, 85\% is accounted for by the theoretical $\chi^2$ distribution. The 15\% in the right tail is mostly attributable to a slightly optimistic noise model designed to avoid underestimating $\chi^2$. This close agreement between predicted and observed $\chi^2$-distributions for the lincal results suggests that except during periods that get automatically flagged, our instrument and analysis pipeline is free from significant systematic errors.
The fully automatic nature of our calibration pipeline and data quality assessment is encouraging for future instruments with data volume too large for direct human intervention.

Calculating $\chi^2$/DoF for flagging and data quality assessment requires an accurate model of noise in the measured visibilities. To compute the noise $\sigma_{ij}$, we approximate $\sigma_{ij}^2$ by $\langle\sigma^2\rangle$, where the average is over all baselines. This assumption that all antennas have the same noise properties drastically deceases the computational cost of calculating $\chi^2$/DoF. Because we have $10^3$ baselines, and the variation of $\sigma_{ij}$ between baselines is less than 20\% (due to slightly different amplifier gains), this approximation should cause only about a 1\% error in the final $\chi^2$/DoF.

To compute $\langle\sigma^2\rangle$, we perform linear regression on each visibility $v_{ij}$ over one minute to obtain its estimated variance $\sigma_{ij}^2$, and then average all $\sigma_{ij}^2$ to obtain $\sigma^2$. Thus we have $\langle\sigma^2\rangle$ at all frequencies every minute. Before we plug $\langle\sigma^2\rangle$ into Equation \ref{eqchisq}, we model it as a smooth and separable function: $\langle\sigma^2\rangle(f,t) = F(f)T(t)$, where $F(f)$ and $T(t)$ are polynomials. The smooth model has three advantages. The first is that it is physically motivated to model thermal fluctuation as a smooth and separable function. Secondly, a smooth noise model makes the $\chi^2$/DoF a much more sensitive flagging device. Theoretically, $\chi^2$/DoF should not rise above 1 when unwanted radio events such as radio frequency interference (RFI) occur,  because they are far field signals that do not violate any redundancy. However, since RFI events make both the signal and noise stronger, by demanding a smooth noise model, the $\langle\sigma^2\rangle$ we use will underestimate the noise during RFI events and give abnormally high $\chi^2$/DoF, which can then be successfully flagged with the $\chi^2/$DoF$\><1.2$ threshold. Thirdly, seasonal changes aside, the noise model is expected to largely repeat itself from day to day, so for future experiments that will operate for years, it suffices to use the same model repeatedly without recomputing $\sigma_{ij}$ {\it in situ} for all the data. Thus, by using a smooth noise model, one can drastically reduce the occurrence of false negatives (since it is better to flag good data than it is to fail to flag bad data) as well as the computational cost of calculating $\chi^2$/DoF.

\subsubsection{Optimal Filtering of Calibration Parameters }\label{secwiener}

While the above-mentioned estimates of the calibration parameters that we obtain from redundant baseline calibration vary over time and frequency, much of that variation is due to the noise in raw data. To minimize the effect of instrumental noise on the calibration parameters, we would like to optimally average information from nearby times and frequencies to estimate the calibration parameters for any particular measurement. 

As we will show below, the optimal method for performing this averaging is Wiener filtering. In the rest of this section, we first measure the  power spectrum of the calibration parameters over time and frequency, and make a determination of how to decompose this into  contributions from signal (true calibration changes) and noise. We  then weight the Fourier components in a way that is informed by their signal-to-noise ratio, and quantify how this Wiener filtering procedure improves upon more naive averaging over time and/or frequency. Finally, we discuss the implications for how regularly (in time and frequency) we should calibrate. It is important to note that while these methods are applied only to MITEoR below, they are applicable to any current or future experiment. 

We model the measured calibration parameter $g_i(f,t)$ for the $i^{\rm th}$ antenna 
as the sum of a true calibration parameter $s_i(f,t)$ (the ``signal'') and uncorrelated noise $n_i(f,t)$:
\begin{equation}
g_i(f,t) = s_i(f,t) + n_i(f,t).
\end{equation}
We choose our estimator $\hat{g_i}(f, t)$ of the true calibration parameter $s_i(f, t)$ to be a linear combination of the observed calibration parameters $g_i$ at different times and frequencies:
\begin{equation}
\hat{{g_i}}(f, t)\equiv\int\int W(f,t,f',t') g_i(f',t') df' dt'
\end{equation}
for some weight function $W$.
We optimize the estimator $\hat{{g_i}}$ by choosing the weight function $W$ that
minimizes the mean-squared estimation error $\langle|\hat{{g_i}}(f, t)-{s_i}(f, t)|^2\rangle$.
Assuming that the statistical properties of the signal and noise fluctuations are stationary over time\footnote{We perform this analysis on data over 12 MHz and two hours, where the signal and noise power are empirically found to be approximately time-independent.}, all correlation functions become diagonal in Fourier space:
\begin{eqnarray}
\langle\tilde{{s}_i}(\tau, \nu)^*\tilde{s}_i(\tau', \nu')\rangle &=& (2\pi)^2 \delta(\tau'-\tau)\delta(\nu'-\nu) S(\tau,\nu),\nonumber\\
\langle\tilde{n}_i(\tau, \nu)^*\tilde{n}_i(\tau', \nu')\rangle &=& (2\pi)^2 \delta(\tau'-\tau)\delta(\nu'-\nu) N(\tau,\nu),\nonumber\\
\langle\tilde{s}_i(\tau, \nu)^*\tilde{n}_i(\tau', \nu')\rangle &=& 0,
\end{eqnarray}
where tildes denote Fourier transforms and $S$ and $N$ are the power spectra of signal and noise, respectively.
This means that the optimal filter becomes a simple multiplication $\hat{\tilde{g}}=\tilde{W}\tilde{g}$
in Fourier space, corresponding to the 
weighting function $\tilde{W}(\tau, \nu)$ that minimizes the mean-squared error 
\begin{equation}
\langle |\tilde{W}(\tau, \nu)\tilde{g}_i(\tau, \nu) - \tilde{{s}_i}(\tau, \nu)|^2\rangle.
\end{equation}
Requiring the derivative of this with respect to $\tilde{W}$ to vanish gives the Wiener filter  \citep{Wiener1942}
\begin{equation}
\tilde{W}(\tau, \nu) = \frac{S(\tau, \nu)}{S(\tau, \nu)+N(\tau, \nu)}.
\end{equation}
\label{WienerEq}
Since $S$ and $N$ are to a reasonable approximation independent of the antenna number $i$, we have dropped all subscripts $i$ for simplicity.
Back in real space, the optimal estimator $\hat{g}_i$ for the $i^{\rm th}$ calibration parameter is thus $g_i$ convolved with 
the 2D inverse Fourier transform of $\tilde{W}$.

 \begin{figure}
\includegraphics[width=80mm]{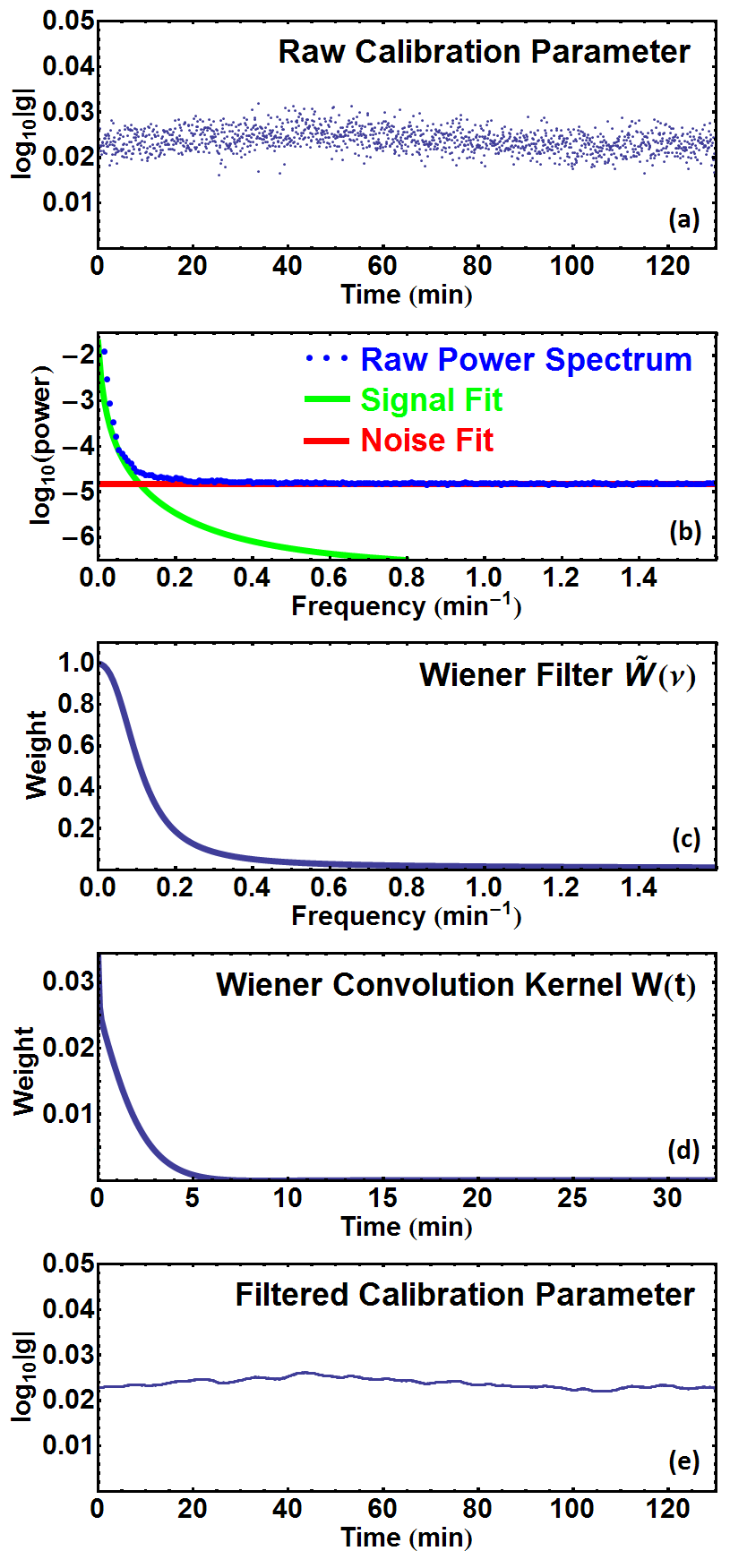}
\caption{
Illustration of 1D Wiener filtering of calibration parameters at different times. Panel (a) shows the amplitude of calibration parameters measured for one antenna over two hours. Panel (b) shows that the average power spectrum across all antennas (blue dots) is well fit by a white noise floor (red horizontal line) plus a sum of two power laws (green curve). Panel (c) shows the Wiener filter in frequency domain computed using Eq. \ref{WienerEq} and the power spectra from panel (b). Panel (d) shows the Wiener convolution kernel in the time domain, the Fourier transform of the filter in Panel (c). Panel (e) shows the best estimates of the true calibration amplitude. The effectiveness of this filter is compared with that of other filters in Table \ref{WienerTab}.
\label{Wiener}
}
\end{figure}
To demonstrate this technique, we show the above process carried out in the time dimension in \fig{Wiener}. In practice we perform the analysis on time and frequency dimensions simultaneously through a 2D FFT. 
The noise power spectrum $N(\nu)$ is seen to be constant to an excellent approximation, corresponding to white noise (uncorrelated noise in each sample). The signal power spectrum $S(\nu)$ is seen to be well fit by a combination of two power laws: 
$S(\nu)\approx (\nu/2.9\times10^{-5}\,\text{Hz})^{-2.7} +(\nu/4.8\times10^{-17}\,\text{Hz})^{-0.46}$.
The optimal convolution kernel $W$ is seen to perform a weighted average of the data on the timescale of roughly 200\,s and frequency scale of 0.15\,MHz, giving the greatest weight to nearby times and frequencies, resulting in an order-of-magnitude noise reduction.

To quantify the effectiveness of the obtained filter compared to  naive ``boxcar'' averages, we use the 2D power spectrum and noise floor of the calibration parameters obtained from real data to simulate many realizations of calibration parameters $g(f,t)=s(f,t)+n(f,t)$, apply various averaging/convolution schemes $W(f,t)$ on the simulated data, and compare their effectiveness by computing the RMS error $\langle |(W\star g)(f,t)-s(f,t)|^2\rangle$ normalized by $\langle{|n|^2}\rangle$. Due to our limited frequency bandwidth as well as frequent RFI contamination, power spectrum modeling in the frequency dimension is very challenging, so the frequency Wiener filter appears to be less effective than the time filter. In Table \ref{WienerTab} we list the normalized noise powers using frequency Wiener filter, time Wiener filter, 2D Wiener filter, as well as traditional boxcar averaging, and the 2D Wiener filter produces results three times less noisy than that of the traditional boxcar averaging.

We have described how to optimally average calibration parameters when we calibrate very regularly in time and frequency. For a future instrument such as an omniscope with $10^6$ antennas, calibration will pose a serious computational challenge, so it is important to know what the minimal frequency one needs to calibrate the instrument. The above analysis conveniently provides an answer to this question. As shown in the second panel of \fig{Wiener}, the signal\footnote{We only show results for amplitude calibration parameters for brevity, as the phase calibration results have nearly identical power spectrum.} is band limited. By the Nyquist theorem, one needs to sample with at least double the frequency of signal bandwidth, so in our case we could measure the calibration parameters without aliasing problems as long as we calibrate once per minute. Calibrating more frequently than this simply helps average down the noise.
Although this one-minute timescale depends on the temporal stability characteristics of the amplifiers and other components used in our particular experiment, it provides a useful lower bound on what to expect from future experiments whose analog chains are even more stable. 

\begin{table}
\begin{tabular}{ | m{4.8cm} | m{2.7cm} | }
	\hline
	\vskip0.5mm
	Averaging method & Relative noise power \\ \hline
	\vskip0.5mm
	No average & 1 \\ 
	\vskip0.5mm
	Frequency Wiener filter & 0.33 \\ 
	\vskip0.5mm
	Time Wiener filter & 0.12 \\ 
	\vskip0.5mm
	Time and frequency Wiener filter & 0.09  \\ 
	\vskip0.5mm
	Time and frequency boxcar average & 0.32 \\ \hline
  \end{tabular}
\caption{
Wiener filtering reduces the noise contribution to the calibration parameters by an order of magnitude. This table lists residual noise power (normalized by original noise power) after applying various filters to average the amplitude calibration parameters in time and/or frequency. The optimal two-dimensional Wiener filter indeed performed the best, lowering the noise power by an order of magnitude. In comparison, the naive boxcar average, using the characteristic scales of the optimal Wiener filter (200\,s and 0.15\,MHz), has more than 3 times residual noise power than the Wiener filtered result.
\label{WienerTab}
}
\end{table}
\subsection{Absolute Calibration}\label{secabscal}
The absolute calibration of the array involves two separate tasks. One is to find the overall gain and to break phase degeneracies  that redundant baseline calibration is unable to resolve, and the other to calibrate fixed properties of the instrument such as the orientation of the array and shape of the primary beam. The former is done by comparing the data to a sky model comprised of the global sky model (GSM)  \citep{GSM} and published astronomical catalogs (see  \citealt{dannyCal} for example).  The latter is done using bright point sources with known positions. While we can take advantage of the extremely high signal-to-noise data in the ORBCOMM channels (around 137~MHz), thanks to the dynamic range provided by our 8 bit correlator, it is important to note that all the algorithms described here are applicable to astronomical point sources as well. 

This section is divided into three parts. The first part describes how we use prior knowledge of the sky to break the degeneracies in redundant calibration results, a vital step to obtain usable data products. The second and third part each describe one aspect of absolute calibration using satellite data: primary beam measurement and array orientation.

\subsubsection{Breaking Degeneracies in Redundant Calibration}\label{secdegeneracy}
Redundant calibration alone cannot produce directly usable data products, due to the degeneracies intrinsic to the algorithms. There is one degeneracy in the amplitude of the calibration coefficients $g_i$, since scaling the amplitude of everything up by a common factor does not violate any redundancy (the degeneracies discussed here are per frequency and per time, as are the calibration solutions). There are three degeneracies in phase, corresponding to three degrees of freedom in a two dimensional linear field (see Appendix \ref{appdegeneracy} for a detailed discussion). In general, breaking these degeneracies requires prior knowledge of the sky. In this section, we briefly describe our algorithm that uses the global sky model (GSM) of  \citet{GSM} to remove these degeneracies. Doing so requires efficiently simulating the response of the instrument to the GSM; we summarize a fast algorithm for doing so in Appendix \ref{appfastgsm}. We defer detailed comparison of our data and the GSM to a future publication.

Our degeneracy removal procedure is an iterative loop that repeats two steps. The first step is to fit for the amplitude degeneracy factor. The knowledge of the GSM and bright point sources give us a set of model visibilities, $m_{ij}^a$, where index $a$ denotes different modeled components such as the GSM or Cygnus A. A linear combination of these models should be able to fit our measurements\footnote{We allow each model to have a separate weight to guard against potential calibration offsets between existing models.}. Thus we fit for the weights $w_a$ of the models by minimizing 
\begin{equation}
\left|v_{ij}-\sum_a w_a m_{ij}^a\right|^2.
\end{equation}
The second step is to break the degeneracy in redundant phase calibration by fitting for the degeneracy vector $\boldsymbol{\mathnormal{\Phi}}$ and the constant $\psi$ defined in Appendix \ref{appdegeneracy}. We assume that the error in the first step's fitting is mostly due to the phase degeneracies, so we take the $w_a$ from step one and fit for $\boldsymbol{\mathnormal{\Phi}}$ and $\psi$ by minimizing
\begin{equation}
\left|\arg(v_{ij})-\arg\left(\sum_a w_a m_{ij}^a\right)-\boldsymbol{d}_{i-j}\cdot\boldsymbol{\mathnormal{\Phi}}-\psi\right|^2,
\end{equation}
where $\boldsymbol{d}_{i-j}$ is the position vector for baseline $i-j$.

\begin{figure*}
\includegraphics[width=180mm]{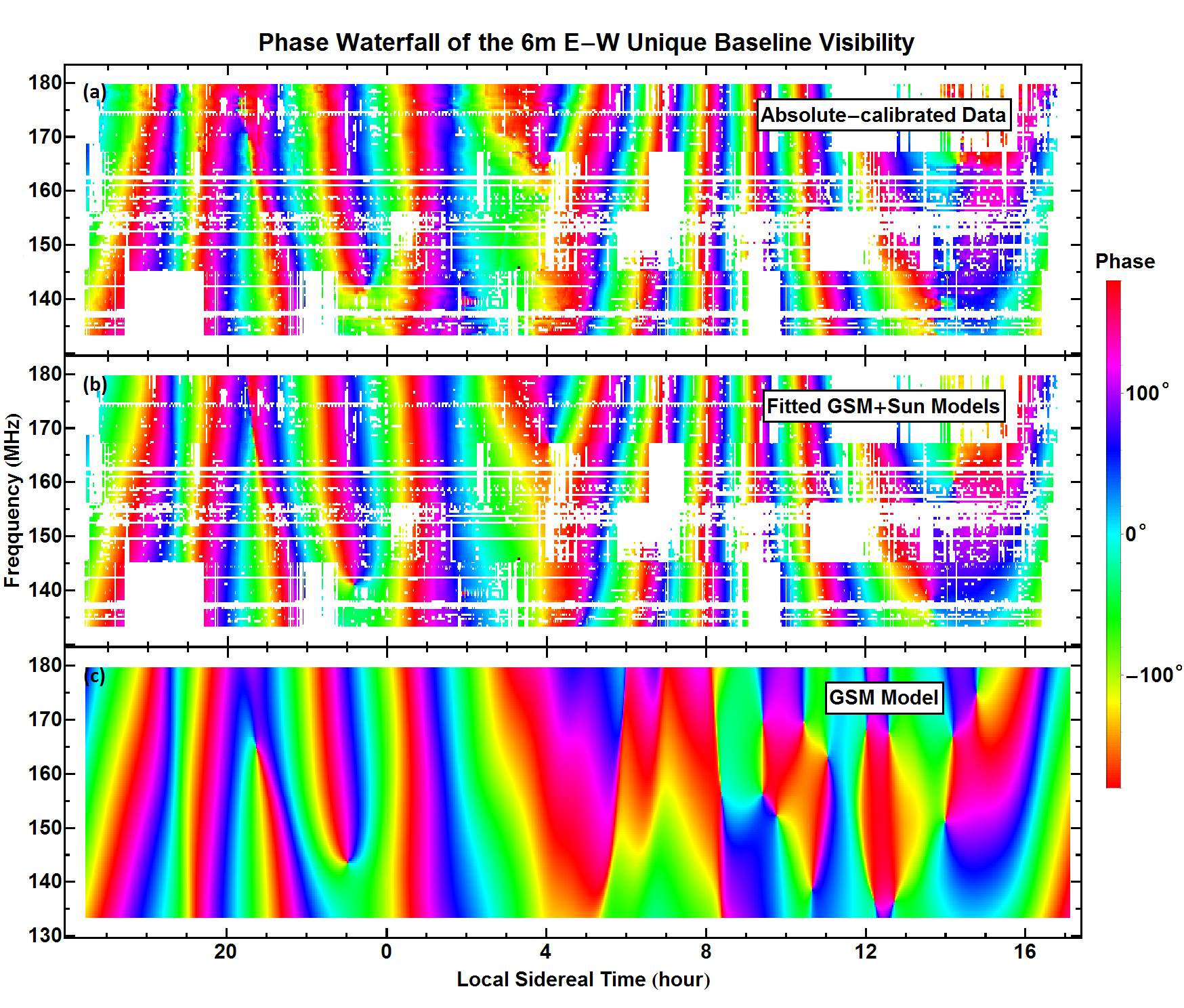}
\caption{
Waterfall plots of phases on the 6~m E-W baseline. These show that our absolute calibration successfully matches the data (panel (a)) with a linear combination of global sky model and known point sources, including the Sun (panel (b)). Panel (c) shows the global sky model alone. The white areas are flagged out using $\chi^2$ criterion described in \fig{chisqwaterfall}. Each plot is stitched together using four independently measured and calibrated frequency bands, aligning local sidereal time. Thus the discontinuities between hours 4 and 12 are due to the Sun rising at different local sidereal times on different days of our observing expedition.
\label{waterfall}
}
\end{figure*}

Note that the two fitting processes described above are not independent of one another, so we repeat these steps until convergence is reached. We find that in practice, the errors converge within two iterations. Our preliminary result is illustrated in \fig{waterfall}, which shows that the data agree very well with current models.

\subsubsection{Beam Measurement Using ORBCOMM Satellites}\label{secbeam}

In general, \textit{in situ} measurements of antenna primary beams over large fields of view pose a challenge to 21\,cm cosmology, as primary beam uncertainties are intimately related to calibration, imaging, and catalog flux uncertainties  \citep{jacobsforeground}. Motivated by these difficulties,  \citet{JonniePrimaryBeam} present a solution that uses celestial point sources and assumes reflection symmetry of the beam, whereas Neben, Bradley, and Hewitt (in preparation) demonstrate high dynamic range beam measurement using the constellation of ORBCOMM satellites. Here, we present \textit{in situ} primary beam measurements of the MWA bow-tie antennas using the ORBCOMM constellation. We take advantage of both the high signal-to-noise ratio of ORBCOMM signals, and of our full cross-correlation measurements (rather than auto-correlations alone) to determine the beam.

In order to measure their primary beam profile $B_{\rm mwa}(\hat{\bf r})$, we compare measurements with MWA antennas to simultaneous measurements with simple center-fed dipoles, whose beam pattern $B_{\rm dipole}$ is known analytically. When there is a single extremely bright point source in the sky, such as an ORBCOMM satellite, we can compute the ratio of the visibilities of select baselines to obtain the ratio of the MWA antenna beam to the analytically known
dipole antenna beam --- thus determining the MWA antenna beam itself. To perform this analysis, two dipole antennas, one orientated along the $x$-polarization axis of the array and the other along the $y$-polarization axis, are added to the array and cross-correlated with all other MWA antennas.

The rationale behind this technique is as follows. At an angular frequency $\omega$, the electric field from a sky signal at the position of a receiving antenna can be encoded in the Jones vector $\boldsymbol{S}(\hat{\boldsymbol{k}})$, where $\hat{\boldsymbol{k}}$ is the position vector of the source \citep{collett2005field}. With a primary beam matrix $\mathbf{B}_j(\hat{\boldsymbol{k}})$, the signal measured by the $j^{\rm th}$ antenna at position $\boldsymbol{r}_j$ is
\begin{equation}
s_j =\int e^{-i [\boldsymbol{k}\cdot \boldsymbol{r}_j+\omega t]}\mathbf{B}_j({\hat{\boldsymbol{k}}})  \boldsymbol{S}(\hat{\boldsymbol{k}}) \; \mathrm{d} \Omega.
\end{equation}
When a single ORBCOMM satellite is above the horizon\footnote{There is typically more than one ORBCOMM satellite above the horizon at any one time, but they are coordinated so that they do not transmit in the same frequency band simultaneously.}, its signal strength is so dominant at its transmit frequency that $\boldsymbol{S}(\hat{\boldsymbol{k}})$ becomes well-approximated by a point source at the satellite's location. The measured signal can then be written as:
\begin{equation}
s_j \approx e^{-i [ \boldsymbol{k}_s \cdot \boldsymbol{r}_j+\omega t]} \mathbf{B}_j(\hat{\boldsymbol{k}}_s) \boldsymbol{S}_s,
\end{equation}
where $\boldsymbol{k}_s$ is the wave vector of the satellite signal, and $\boldsymbol{S}_s$ is the Jones vector encoding the satellite signal strength.

If we limit our attention to either $x$-polarization or $y$-polarization and approximate the off diagonal terms of $\mathbf{B}(\hat{\boldsymbol{k}})$ as zero, the visibility for two antennas can be written as
\begin{equation}
v_{jk} \approx S^2 B_j(\hat{\boldsymbol{k}}_s)^* B_k(\hat{\boldsymbol{k}}_s)e^{-i \boldsymbol{k}_s \cdot(\boldsymbol{r}_k-\boldsymbol{r}_j)}.
\end{equation}
If we take one visibility $v_{ij}$ formed by correlating a simple center-fed dipole with an MWA bow-tie antenna and another visibility $v_{kl}$ for the same baseline vector formed by correlating two MWA antennas, 
then their ratio is simply
\begin{equation}
 \frac{|v_{ij}|}{|v_{kl}|} \approx \frac{|B_{\rm mwa}(\hat{\boldsymbol{k}}_s)|}{|B_{\rm dipole}(\hat{\boldsymbol{k}}_s)|},
 \label{Eq.correlation ratio}
\end{equation}
because the satellite intensity $S$ and one MWA beam factor $B_{\rm mwa}$ all cancel out, and the phase factor $e^{-i \boldsymbol{k}_s \cdot(\boldsymbol{r}_j-\boldsymbol{r}_i)}$ is removed due to taking absolute values of the visibilities. This means that when a single point source dominates the sky, the ratio of visibility amplitudes is simply the ratio of the antenna beams at the direction of the point source. Since we already know the beam 
$B_{\rm dipole}$ of a center-fed dipole over a ground screen, we can directly infer the magnitude of MWA primary beam $|B_{\rm mwa}(\hat{\boldsymbol{k}}_s)|$.

\begin{figure}
\centerline{\includegraphics[width=85mm]{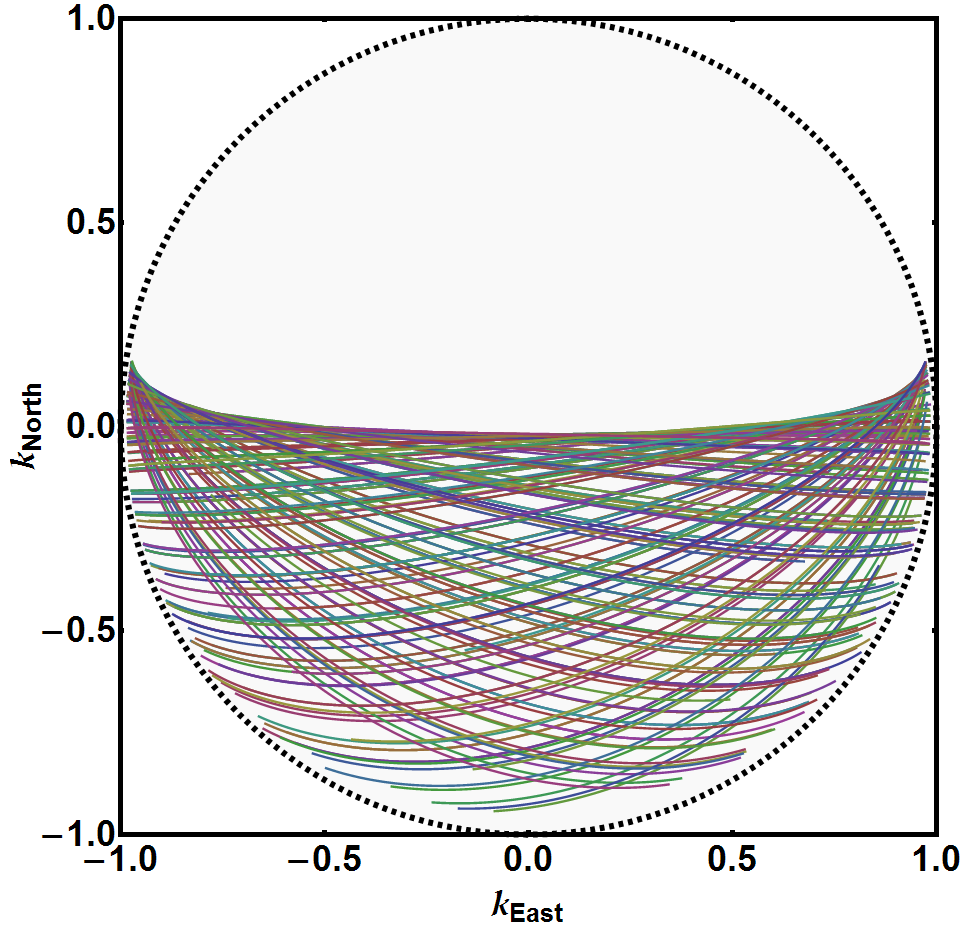}}
\caption{Projected trajectories of 248 passes of ORBCOMM satellites over 40 hours. With these passes we obtain sufficiently dense sampling of the MWA antennas primary beam that we can robustly map its response, especially at high elevations where the response is strongest. With a map of the southern half of the primary beam, we can use the reflection symmetry of the antennas to infer the entire beam at the ORBCOMM transmission frequencies. Each curve is a satellite pass projected onto the x-y plane, and the different colors specify sets of data taken at different times.}
\label{fig:sat passes}
\end{figure}

In order to fully map out the MWA primary beam, we need to take data during many satellite passes until we have direction vectors that densely cover the entire sky. Satellite signals from 27 ORBCOMM satellites at 5 frequencies in the range of 137.2-137.8 MHz were identified.  Their orbital elements are publicly available\footnote{We obtained the TLE files from CelesTrak, a company that archives TLEs of many civil satellites.}, so we can calculate $\hat{\boldsymbol{k}}_s(t)$ straightforwardly. With 40 hours of data taken at the frequencies of interest, we were able to obtain 248  satellite passes, shown in \fig{fig:sat passes}.

\begin{figure}
\includegraphics[width=85mm]{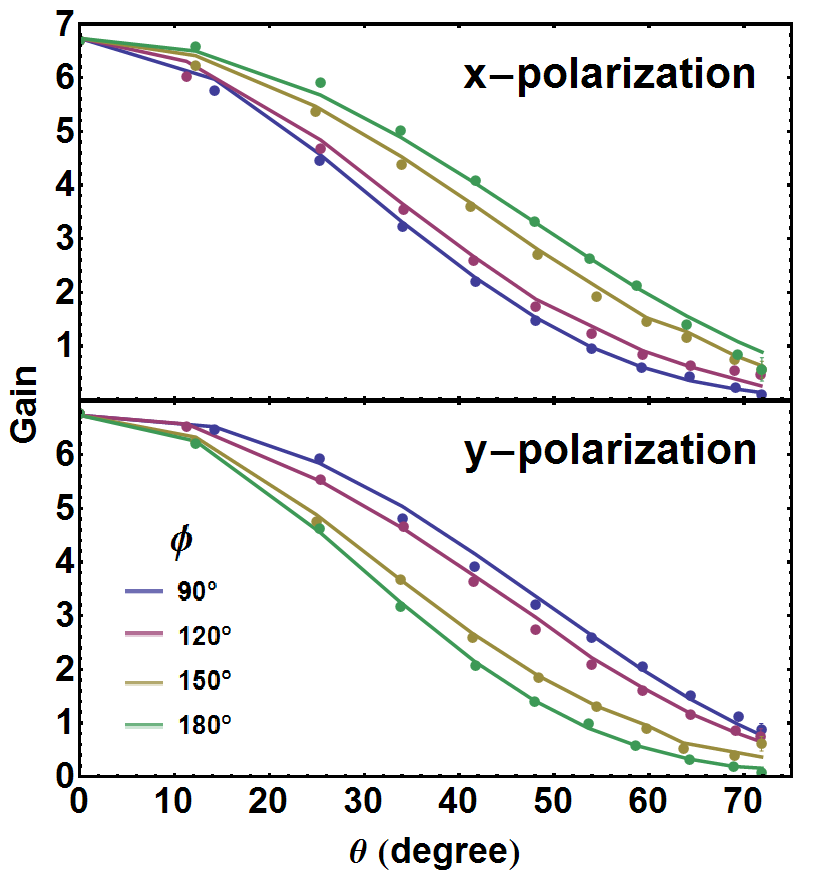}
\caption{Measured MWA primary beam patterns compared to those obtained from numerical modeling. The two panels show the predictions (curves) and measurements (points) of the primary beam for the $x$-polarized and $y$-polarized MWA antennas. Each curve shows how the primary beam changes with the polar angle $\theta$ for a fixed
azimuth angle $\phi$. To reduce noise, the measurements have been averaged in 10 square degree bins.}
\label{fig:azimuth slice}
\end{figure}

We compared our measurements of the MWA primary beam using Equation \ref{Eq.correlation ratio} to numerical calculations using the FEKO electromagnetic modeling software package. Fixing an azimuth angle $\phi$, we can plot and compare the simulated and measured beam at different polar angles $\theta$ (the angle between the direction vector and zenith). \fig{fig:azimuth slice} shows how the beam changes with $\theta$ for four different $\phi$-values, where $\phi=0$ correspond to North and increases clockwise.  Our measurements of the MWA beams are seen to agree well with the numerical predictions for both polarizations. The small differences between the predicted and measured beams are larger than the statistical noise, implying that the main limitation is not noise but one or more of the above-mentioned approximations, or approximations in the electromagnetic antenna modeling.

\subsubsection{Calibrating Array Orientation Using ORBCOMM Satellites and the Sun}\label{secrotation}
\begin{figure}
\hskip-1.5mm\centerline{\includegraphics[width=85mm]{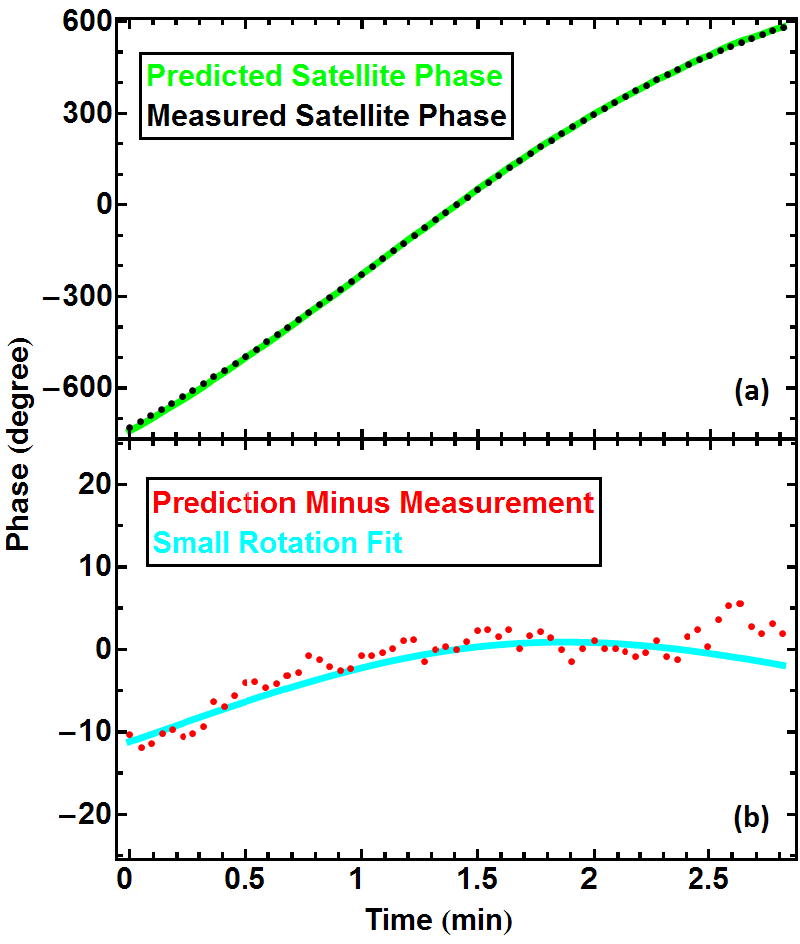}}
\caption{
Illustration of using the high signal-to-noise ORBCOMM data to calculate any small rotation in the array relative to the field-measured orientation. Panel (a) shows the rapidly wrapping phase of the raw data (black) from one baseline at the ORBCOMM frequency during the peak three minutes of a single satellite pass. In green, we see the predicted values computed with the field-measured array orientation and publicly available satellite positions. The residual between the model and the data is plotted in red points in panel (b). Finally, the cyan curve shows the best fit using small angle rotations of the array. In practice we use hundreds of satellite passes and all the baselines to obtain a single accurate fit for the true orientation of the array.
}\label{sat}
\end{figure}

The orientation of the array is very important, because the degeneracy removal process relies on the predicted measurement for each unique baseline, which in turn relies on precise knowledge of the baselines' orientations. Although we measured the relative position of each antenna to millimeter level precision with a laser-ranging total station, we did not measure the absolute orientation of the array to better than the $\sim 1^\circ$ accuracy obtainable with a handheld compass. To improve upon this crude measurement, we make use of both the known positions of both ORBCOMM satellites and the Sun. As we show in \fig{sat}, the exceptional signal-to-noise in the ORBCOMM data allows us to fit for a small array rotation as a first order correction to a model based on our crude measurement.
Our method for finding the true orientation of the array is as follows. For a given baseline during the peak few minutes of an ORBCOMM satellite pass at frequency $\nu$, we measure a phase $\phi(t)$. We also know the satellite's position vector $\boldsymbol{k}(t)$.  However, we only have crude knowledge of baseline vector $\boldsymbol{d}_0$ in units of wavelength, where vectors are in horizontal coordinates with $x,y,z$ that correspond to south, east and up. We can therefore only compute a crude prediction of the phase measurement 
\begin{equation}
\phi_0(t) = 2\pi \boldsymbol{k}(t)\cdot\boldsymbol{d}_0.
\end{equation}
We assume that the difference between the measurement $\phi(t)$ and our crude prediction $\phi_0(t)$ is due to a small angle rotation of the baseline vector $\boldsymbol{d}_0$ around the axis $\pmb{\theta}=(\theta_x,\theta_y,\theta_z)$ by an angle $\theta = |\pmb{\theta}|$, ignoring a constant cable length delay.\footnote{Here it is important to use data before redundant baseline calibration to avoid phase degeneracy. We remove the phase delay from cables by allowing a constant offset that matches $\phi(t_M)$ with the crude prediction at time $t_M$ when the satellite has the strongest signal during the pass.} In the small $\theta$ regime, we have that
\begin{align}
\phi(t) - \phi_0(t) 
&= 2\pi \boldsymbol{k}(t)\cdot(\mathbf{R}(\pmb{\theta})\cdot\boldsymbol{d}_0 - \boldsymbol{d}_0) \nonumber \\
&\approx 2\pi \boldsymbol{k}(t)\cdot(\pmb{\theta}\times\boldsymbol{d}_0) \nonumber \\
&= 2\pi (\boldsymbol{d}_0\times\boldsymbol{k}(t))\cdot\pmb{\theta},
\end{align}
where $\mathbf{R}(\pmb{\theta})$ is the rotation matrix.
Since we have a set of equations each representing a different time, the problem of finding $\pmb{\theta}$ can be reduced to that of finding a least squares fit.
With 117 satellite passes, we obtained the following best fit for the array rotation around the vertical axis:
$$\theta_z^\text{sat}=0.66^{\circ}\pm0.0005^{\circ}_{\text{stat}}\pm0.07^{\circ}_{\text{sys}}.$$

While this method is very precise for solving the main problem we were worried about---the direction of North ($\theta_z$) which we approximated in the field with a handheld compass---it is less useful for measuring rotations of the array in the other two directions. Our instrument's absolute timing precision is only $\sim 0.5$ seconds, which makes it hard to distinguish rotations about the North-South axis from timing errors, as most ORBCOMM passes are East-West. This issue can of course be easily addressed in future experiments; for our experiment, we solve it using a more slowly moving bright point source: the Sun.

By using one day of solar data at 139.3\,MHz, we obtained
\begin{align*}
(\theta_x,\theta_y,\theta_z)^{\odot} =&(-0.08^{\circ}, -0.12^{\circ}, 0.672^{\circ})\\
\pm&(0.01^{\circ}, 0.03^{\circ}, 0.004^{\circ})_{\text{stat}}\\
\pm&(0.04^{\circ}, 0.003^{\circ}, 0.005^{\circ})_{\text{sys}}.
\end{align*}
Although solar data is noisier, in part because the Sun is not as bright as the ORBCOMM satellites in a given channel, timing errors are no longer important. These results agree with and complement the satellite-based results and allow us to confidently pin down the orientation of the array and thus improve the quality of the calibration of all of our data. The excellent agreement between the independent measurements 
$\theta_z^{\rm sat}\approx 0.66^\circ$ and
$\theta_z^{\odot}\approx 0.67^\circ$ 
provides encouraging validation of both the satellite and solar calibration techniques.

\subsection{Systematics}\label{secsystematic}
As we discussed in section \ref{secchisq}, most of our data are well-calibrated with $\chi^2/\mbox{DoF} <1.2$, which means that any systematic effects should lie well below the level of the thermal noise. In this section we aim to identify all the systematic effects present in the system, and describe our efforts to quantify and, whenever possible, remove them. The systematics can be categorized into two groups: 
\begin{enumerate}
\item Signal-dependent systematics that grow as the signal becomes stronger, such as cross-talk, antenna position errors and antenna orientation errors.
\item Signal-independent systematics, such as radio frequency interference (RFI) from outside or inside the instrument.
\end{enumerate}
Below we find a strict upper bound of $0.15\%$ for the signal-dependent component, as well as a signal-independent component which is easy to remove. 

\begin{figure}
\centerline{\includegraphics[width=88mm]{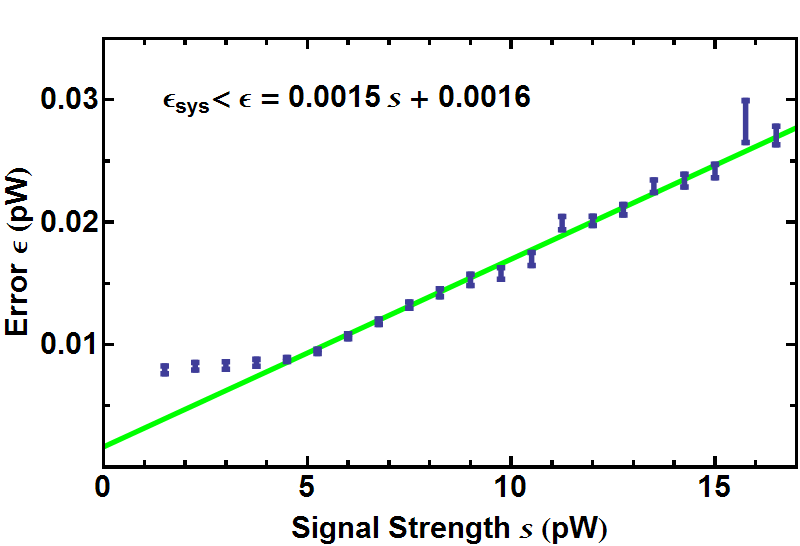}}
\caption{
Signal-dependent systematic error and its linear fit. By comparing the modeled and calibrated visibilities during ORBCOMM satellite passes, we conclude that signal-dependent systematic effects account for no more than $0.15\%$ of our measurement. We calculate the average fitting error per baseline $\epsilon = \langle|v_{ij}-y_{i-j}g_i^*g_j|\rangle$ and the average signal strength $s = \langle|y_{i-j}|\rangle$ binned over one day's data (blue points). The green line fits the data points above the noise floor. While many systematic errors, such as cross-talk, can contribute to the fitting error in addition to thermal noise, the best-fit slope of 0.0015 puts an upper bound on the sum of all signal-dependent errors. Since the ORBCOMM signal is so strong, any signal-independent systematic errors are negligible in this analysis. The high noise floor of $\sim 0.01$~pW is due to our digital tuning in the ORBCOMM frequency channels to maximize dynamic range.}\label{saterr}
\end{figure}

To quantify signal-dependent systematics, we again use ORBCOMM satellite data. Because the ORBCOMM signals are $10^3$ times brighter than astronomical signals, and we know that any signal-independent systematics must be weaker than the astronomical signals (otherwise they would have been blatantly apparent in the data), any signal-independent systematics must be negligible compared to the ORBCOMM signal. We therefore investigate how the discrepancies between calibrated visibilities and the models for each unique baseline depend on ORBCOMM signal strength.  We define the average fitting error per baseline at a given time and frequency to be
\begin{equation}
\epsilon = \langle|v_{ij}-y_{i-j}g_i^*g_j|\rangle,
\end{equation} 
which is a combination of antenna noise and systematic errors. If we compute $\epsilon$ at different times with different signal strength and compute its signal dependence, we can derive an upper bound on the signal dependence of systematic errors. To do this, we take all data at the ORBCOMM satellite frequency over a day and compute $\epsilon$ after performing redundant calibration. We then bin the $\epsilon$-values according to the average signal strength $s = \langle|y_{i-j}|\rangle$, and obtain the results shown in 
\fig{saterr}\footnote{Another way of describing these data points is that, if we look at the third panel in \fig{omniviewer}, we are plotting the average small spread in each unique baseline group versus the radius of the circle, and as the satellites pass over, both the circle size and the amount of average spread change over time, forming the data set in question.}.
The result is seen to be well fit by a constant noise floor plus a 
straight line $\epsilon\approx 0.0015 s$.
This slope implies that the combined effect of all signal-dependent systematic effects is 
at most 0.15\%.
This is merely an upper bound on the systematics, since it is possible that the increase in $\epsilon$ is mainly due not to systematics but to an increase in instrumental noise caused by an increase in the system temperature during the ORBCOMM passes.

There is one signal-dependent systematic that is not included in the above analysis: deviation from redundancy caused by imperfect positioning of antenna elements. This is because the data we used to derive the upper bound is always dominated by a single point source, the ORBCOMM satellite, and redundant calibration cannot detect any deviation of antenna position when the sky is dominated by a single point source\footnote{This is because for any arbitrary position deviation $\Delta \boldsymbol{r}_i$ for antenna $i$, one can add a phase equal to $\boldsymbol{k}\cdot\Delta \boldsymbol{r}_i$ to the calibration parameter $g_i$ to perfectly ``mask" this deviation. Note that this ``mask phase" depends on $\boldsymbol{k}$ and thus changes rapidly over time when the ORBCOMM satellite moves across the sky.}. We have two ways to quantify the error in our data due to antenna position errors. Firstly, the laser-ranging measurements of antenna positions in the field indicate an average of 0.037\,m deviation from perfect redundancy, which translates to about 2\% average error in phase on each visibility. Since the deviations are in random directions, the variance of phase error in the unique baseline fits should be brought down by a factor equal to the number of redundant baselines, resulting in phase errors much less than 1\% for most of the unique baselines. Secondly, although satellite calibration cannot detect position error in a given snapshot, over time the position errors would create very rapidly changing calibration parameters, which we do not observe in our data. Lastly, a formalism exists  \citep{omnical} to treat errors in antenna placement as small perturbations when redundantly  calibrating, although though we did not need to take advantage of this technique for the present paper.

We first identified a signal-independent systematic when we obtained consistent $\chi^2/\text{DoF}\sim 4$ for much of our data\footnote{This was before we obtained a consistent $\chi^2/\text{DoF}\sim 1$ in Section \ref{secchisq}, which occurred after we were able to remove the systematic described here.}, which means that the fitting error was on average twice as large as the thermal noise in each visibility. This implies a systematic (or a combination of systematics) at the level of $10^{-6}$\,pW/kHz, about 10\% of the total astronomical signal. Given the above analysis, we can exclude the possibility of any signal-dependent explanations such as cross-talk between channels or antennas. While we are unable to offer any conclusive explanation of this systematic, it appears consistent with persistent near-field RFI, perhaps originating from our electronics. Fortunately, we found this additive signal to vary only very slowly over time, typically remaining roughly constant over 5-minute periods, which made it easy to remove. After calibrating the data with logcal, we average the fitting errors $\epsilon_{ij} = \langle v_{ij}-y_{i-j}g_i^*g_j\rangle_t$ over time and subtract them from the data before we run logcal again. We perform the averaging over 5 minute segments, corresponding to 112 independent time samples, and iterate the calibration-subtraction process three times. This corresponds to less than a 1\% increase in the number of effective calibration parameters we fit for during logcal. Because many baselines probe the same unique baseline, the procedure described above exploits the redundancy of the array to robustly remove this slowly varying, signal-independent systematic, leaving us with $\chi^2/\text{DoF}\sim 1$.

\section{Summary and Outlook}\label{secsummary}

We have described the MITEoR experiment, a pathfinder ``omniscope" radio interferometer with 64 dual-polarization antennas in a highly redundant configuration. We have demonstrated a real-time precision calibration pipeline with automatic data quality monitoring that uses $\chi^2$/DoF as a data quality metric to ensure that redundant baselines are truly seeing the same sky. We have also implemented various instrumental calibration techniques that utilize the ORBCOMM constellation of satellites to measure the primary beam and precise orientation of the array. Our success bodes well for future attempts to perform such calibration in real-time instead of in post-processing, and thus clears the way for FFT correlation that will make interferometers with $\gtrsim 10^3$ antennas cost-efficient by reducing the computational cost of correlating $N$ antennas from an $N^2$ scaling to an $N\log N$ scaling.  It also suggests that the extreme calibration precision required to reap the full potential of 21~cm cosmology is within reach. 

The various calibration techniques that MITEoR successfully demonstrates are now being incorporated into the much more ambitious HERA project\footnote{\url{http://reionization.org/}}  \citep{poberliudillon}, a broad-based collaboration among US radio astronomers from the PAPER, MWA, and MITEoR experiments. Our results are also pertinent to the design of the SKA low-frequency aperture array\footnote{\url{http://skatelescope.org/}}. HERA plans to deploy around 331 14-meter dishes in a close-packed hexagonal array in South Africa, giving a collecting area of more than 0.05 km$^2$, virtually guaranteeing not only a solid detection of the elusive cosmological 21 signal but also interesting new clues about our cosmos. 

{\bf Acknowledgments:}
MITEoR was supported by NSF grants AST-0908848 and AST-1105835, the MIT Kavli Instrumentation Fund, the MIT undergraduate research opportunity (UROP) program, FPGA donations from XILINX, and by generous donations from Jonathan Rothberg and an anonymous donor. We wish to thank Jacqueline Hewitt and Aaron Parsons for helpful comments and suggestions, Dan Werthimer and the CASPER group for developing and sharing their hardware and teaching us how to use it, Matias Zaldarriaga for great help in getting this project started, Dr. T.S. Kelso and CelesTrak for providing TLE files for ORBCOMM satellites, Tony Hopko from ORBCOMM, Jonathon Kocz from Harvard University, Rick Raffanti from Techne Instruments, Stuart Rumley from Valon Technology, and Kyle Slightom from Analog Devices for their expert hardware advice, Jean Papagianopoulos, Thea Paneth, Steve Burns and his family for invaluable help, Meia Chita-Tegmark, Philip and Alexander Tegmark and Sherry Sun for deployment assistance,
and Joe Christopher and his fellow denizens of The Forks for their awesome hospitality and support.

\bibliographystyle{mn2e}
\bibliography{MITEOR3}
\newpage
\appendix

\section{Phase Degeneracy in Redundant Calibration}\label{appdegeneracy}
Both of our redundant baseline calibration algorithms, logcal and lincal (see Section \ref{secloglin}), have the same set of phase degeneracies that require additional absolute calibration that must incorporate knowledge of the sky. When calibrating a given unique baseline, the quantity that logcal minimizes is
\begin{align}\label{eqlogdegen}
\sum_{jk} |(\theta_{j-k}-\phi_j+\phi_k)- \arg(v_{jk})|^2,
\end{align}
where we define $\theta_{j-k} \equiv \arg(y_{j-k}), \phi_j\equiv\arg(g_j)$. Similarly, lincal minimizes
\begin{align}\label{eqlindegen}
&\sum_{jk} |(y_{j-k}g_j^*g_k)-v_{jk}|^2\nonumber\\
=& \sum_{jk} \left||y_{j-k}g_j^*g_k|\exp\left[i(\theta_{j-k}-\phi_j+\phi_k)\right]-v_{jk}\right|^2.
\end{align}

Unfortunately, for all values of $\theta_{j-k}$ and $\phi_k$, one can add any linear field $\boldsymbol{\mathnormal{\Phi}}\cdot\boldsymbol{r}_j+\psi$ to the $\phi_j$ across the entire array while subtracting $\boldsymbol{\mathnormal{\Phi}}\cdot\boldsymbol{d}_j$ from the $\theta_{j-k}$ without changing the minimized quantities:
\begin{align}
 \theta'_{j-k}-\phi'_j+\phi'_k \equiv& ( \theta_{j-k} - \boldsymbol{\mathnormal{\Phi}}\cdot\boldsymbol{d}_{j-k}) -(\phi_j+\boldsymbol{\mathnormal{\Phi}}\cdot\boldsymbol{r}_j+\psi) \nonumber \\ 
&+(\phi_k+\boldsymbol{\mathnormal{\Phi}}\cdot\boldsymbol{r}_k+\psi) \nonumber \\ 
=& \theta_{j-k}-\phi_j+\phi_k.
\end{align}
Here $\boldsymbol{r}_j$ is the position vector of antenna $j$ and $\boldsymbol{d}_{j-k} \equiv \boldsymbol{r}_k-\boldsymbol{r}_j$ is the baseline vector for the unique baseline with best-fit visibility $y_{j-k}$. Thus, the quantities in expressions \ref{eqlogdegen} and \ref{eqlindegen} that the calibrations minimize are degenerate with changes to the linear phase field $\boldsymbol{\mathnormal{\Phi}}$ and the scalar $\psi$. This means that there are, in general, 4 degenerate phase parameters that need absolute calibration: one overall phase $\psi$ and three related to the three degrees of freedom of the linear function $\boldsymbol{\mathnormal{\Phi}}$ (which reduces to two for a planar array).

In an ideal instrument, the measured visibilities for a given unique baseline would be
\begin{align}
y_{i-j} = \int_{k_x,k_y}e^{i\boldsymbol{k}\cdot\boldsymbol{d}_{i-j}}S_B(k_x,k_y)dk_xdk_y,
\end{align}
where $\boldsymbol{k}=(k_x,k_y,k_z)$ is the wave vector of incoming radiation and $S_B(k_x,k_y)$ is the product of the incoming signal intensity and the primary beam in the direction $\hat{\boldsymbol{k}}$ normalized by $k k_z$ (which comes from the Jacobian of the coordinate transformation from spherical coordinates; see  \citealt{FFTT}). When the array is coplanar, we can take an inverse Fourier transform of $y_{i-j}$ and obtain an image of $S_B(k_x,k_y)$.  Above we saw that the best fit $y_{i-j}$ computed by logcal and lincal is multiplied by an unknown linearly varying phase $\boldsymbol{\mathnormal{\Phi}}\cdot\boldsymbol{d}_{i-j}$. 
Since multiplication in $uv$ space is a convolution in image space, this means that the image generated using these $y_{i-j}$ is the true image convolved with a Dirac delta function centered at $\boldsymbol{\mathnormal{\Phi}}$, which corresponds to a simple shift by the unknown vector 
$\boldsymbol{\mathnormal{\Phi}}$ in the $S_B(k_x,k_y)$ image space. 

\begin{figure}
\centerline{\includegraphics[width=85mm]{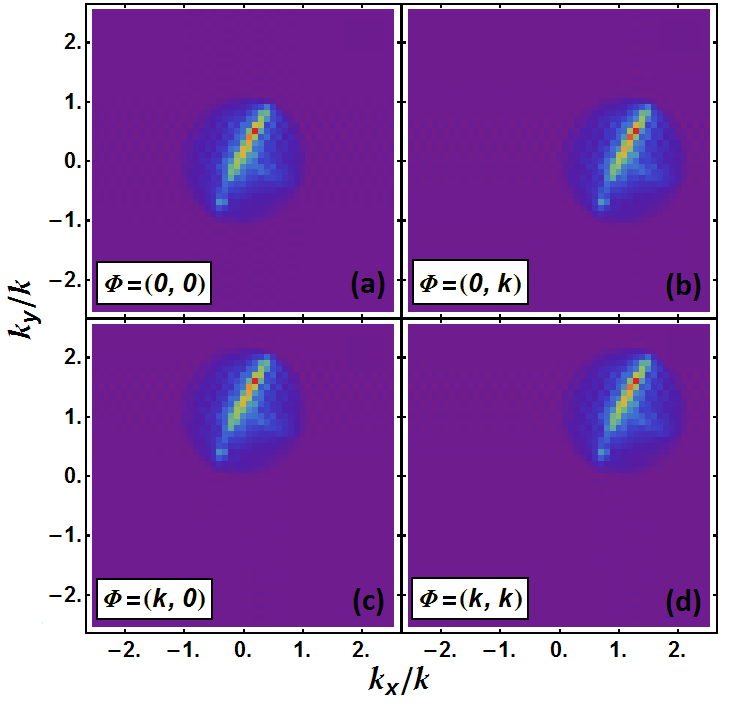}}
\caption{
Illustration of phase degeneracies shifting the sky image where the sky disk is demarcated. The linear phase degeneracy, which takes the form $\boldsymbol{\mathnormal{\Phi}}\cdot\boldsymbol{r}_i$ in each antenna for any $\boldsymbol{\mathnormal{\Phi}}$, corresponds to a shift of the reconstructed image.
These simulated images demonstrate shifts of fiducial sky image at 160~MHz caused by four different $\boldsymbol{\mathnormal{\Phi}}$, where the fiducial array's shortest baseline is 0.2~m.  Panel (a) shows the image obtained from visibilities with no $\boldsymbol{\mathnormal{\Phi}}$, and the sky image is centered at 0. In the other three panels, the sky image is shifted by amount $\boldsymbol{\mathnormal{\Phi}}$. Even if one has no knowledge of what the true sky is, it is still possible to determine 
$\boldsymbol{\mathnormal{\Phi}}$ from where the sky image is centered.
\label{shifts}
}
\end{figure}

To calibrate these last few overall phase factors, one can either make sure that bright radio sources line up properly in the image, or match phases between measured visibilities and predicted visibilities, as we described in Section \ref{secdegeneracy}. However, there may be another complementary way to remove this phase degeneracy without any reference to the sky. We know that physically the true image $S_B(k_x,k_y)$ is only non-zero within the disk
$|k_x^2+k_y^2|^{1/2}\le k$ centered around the origin, and a shift caused by $\boldsymbol{\mathnormal{\Phi}}$ would move this circle off center. This suggests that we should be able to reverse engineer $\boldsymbol{\mathnormal{\Phi}}$ by looking at how much the image circle has been shifted, without knowing what $S_B(k_x,k_y)$ is. \fig{shifts} demonstrates how the image is shifted by $\boldsymbol{\mathnormal{\Phi}}$ using simulated data.

\begin{figure}
\includegraphics[width=87mm]{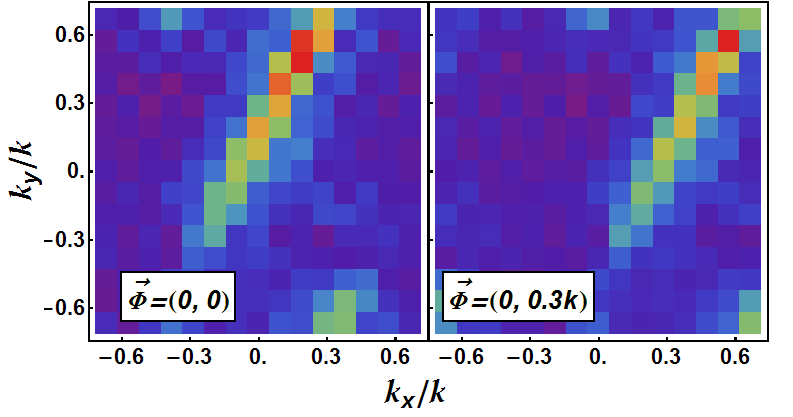} 
\caption{
Illustration of phase degeneracies shifting the sky image where the sky disk is not demarcated. With any practical array configuration, including that of MITEoR, distinguishing image shifts caused by the $\boldsymbol{\mathnormal{\Phi}}$-degeneracy becomes significantly more difficult. 
These images demonstrate shifts of fiducial sky image at 160~MHz just as in \fig{shifts}, but with MITEoR's compact configuration where the shortest baseline is 1.5~m.  In the left panel, the image is obtained from visibilities with $\boldsymbol{\mathnormal{\Phi}}=(0,0)$, and in the right panel the sky image is shifted by and amount $\boldsymbol{\mathnormal{\Phi}}=(0,0.3k)$. Because the shortest baseline is too long ($0.8\lambda$), the Fourier transform of the visibilities only cover up to about 0.7 in $k_x$ and $k_y$, so in contrast with \fig{shifts}, it is impossible to determine $\boldsymbol{\mathnormal{\Phi}}$ by merely looking at where the sky image is centered without prior knowledge of the sky.
\label{shiftsmiteor}
}
\end{figure}

Unfortunately, this simple approach to identifying and removing the effect of $\boldsymbol{\mathnormal{\Phi}}$ suffers from a few complications. By far the most important one is the requirement of very short baselines. In the example in \fig{shifts}, the shortest separation between antennas is $0.21\lambda$, and it is easy to show that the sky disk is only clearly demarcated when the shortest separation is less than $0.5\lambda$\footnote{This is the 2D imaging counterpart of the well-known fact that, in signal processing, one must sample with a time interval shorter than $0.5\nu^{-1}$ to avoid aliasing in the spectrum of maximum frequency $\nu$.}. This sets a limit on the physical size of each element, which makes achieving a given collecting area proportionately more difficult. As \fig{shiftsmiteor} shows, the deployed configuration of MITEoR cannot be used to reverse engineer the degeneracy vector $\boldsymbol{\mathnormal{\Phi}}$ without knowledge of the true sky.

\section{A Hierarchical Redundant Calibration Scheme with $\mathcal{O}(N)$ Scaling}\label{apphierarchical}
One of the major advantages of an omniscope is its $N\log N$ cost scaling where $N$ is the number of antennas. However, existing calibration techniques, including the ones presented in this paper, require all of the visibilities to compute the calibration parameters. Since the cost for computing the visibilities alone scales as $N^2$, this is a lower bound to the computational cost of existing calibration schemes regardless of the actual algorithm. While current instruments with less than $10^3$ elements can afford full $N^2$ cross-correlation, such computation will be extremely demanding for a future omniscope with $10^4$ or more elements. Thus, to take advantage of the $N\log N$ scaling of an omniscope with large $N$, it is necessary to have a calibration method whose cost scaling is less than $N\log N$. In this section, we describe a such a method using a hierarchical approach, and show that its computational cost scales only linearly with the number of antennas. 

\begin{figure}
\centerline{\includegraphics[width=85mm]{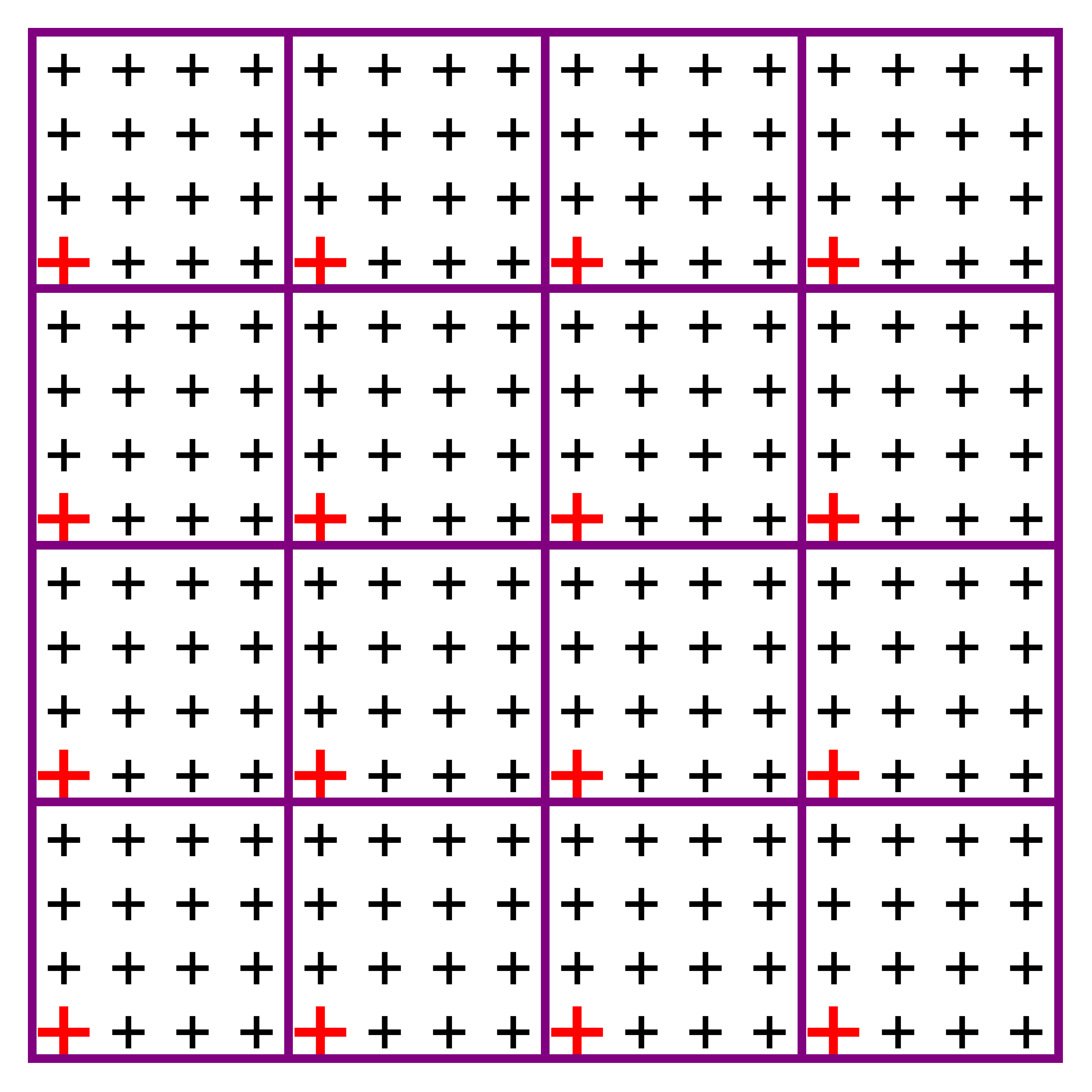}}
\caption{
Example of the hierarchical calibration method for 256 antennas (marked by $+$-symbols) 
viewed  as a 2-level hierarchy of $4\times 4$ arrays $(m=16, n=2)$.
Our method first calibrates each sub-array independently with both relative and absolute calibrations. This produces calibration parameters for every antenna, up to one phase degeneracy $\psi$ per sub-array. We can remove these 16 phase degeneracies among sub-arrays by choosing one antenna out of each sub-array (marked red and thick) and performing calibration on these 16 antennas. Thus we have calibrated the whole 256 antenna array by performing 16-antenna calibration 16+1=17 times. This can be generalized to a hierarchy with more levels by placing 16 such 256-antenna arrays in a $4\times 4$ grid to get a 4096-antenna array, and then repeating to obtain arrays of exponentially increasing size. As shown in the text, the computational cost for this calibration method scales only linearly with the number of antennas.
\label{hierarray}
}
\end{figure}

\Fig{hierarray} illustrates the hierarchical calibration method for an example with a 
256 antennas in a $16\times 16$ regular grid, viewed as a 2-level hierarchy of $4\times 4$ grids.
More generally, consider an $n$-level hierarchy with $m$ sub-arrays at each level, containing a total
of $N=m^n$ antennas; the example in \fig{hierarray} corresponds to $m=16$, $n=2$ and $N=256$.\footnote{It is easy to see that for a regular grid of $N$ antennas, $N$ need not an exact power of $m$ to obtain the scaling that we will derive.}
Let $B_m$ denote the computational cost of calibrating a basic $m$-antenna array\footnote{$B_m$ includes the cost to compute cross-correlations between the $m$ antennas, as well as both relative and absolute calibrations. The cost $B_m$ is unimportant for the scaling as long as it is independent of $n$.}
Let $C_n$ denote the computational cost of calibrating the entire $n$-level hierarchy containing all $N$ antennas.
We have $C_1=B_m$ by definition and 
\begin{equation}
C_{n+1}=m C_n + B_m
\end{equation}
since, as explained in the caption of \Fig{hierarray}, we can calibrate the $m$ sub-arrays at cost $C_n$ each and then calibrate the $m$ reference antennas (one from each sub-array) at cost $B_m$.
Solving this recursion relation gives 
\begin{eqnarray} 
C_n&=&B_m(1+m(1+m(1+m(1+\dotsi))))\nonumber\\
&=&B_m\sum_{k=0}^{n-1}{m^k}
=\frac{m^n - 1}{m-1}B_m\nonumber\\
&=&\frac{N-1}{m-1}B_m =\mathcal{O}(N).
\label{eq:hierarchy}
\end{eqnarray}

Eq.~\ref{eq:hierarchy} implies that for a fixed $m$, the computational cost for calibrating a $10^5$ antenna array will be 10 times that of a $10^4$ antenna array. Intuitively, the cost reduction comes from not computing cross-correlations among most pairs of antennas. In the simple case in \fig{hierarray}, only one visibility is computed between each pair of sub-arrays, rather than 256 visibilities in a full correlation scheme. Because of the reduced number of cross-correlations computed, we expect the quality of calibration parameters to be worse than that in the full correlation case. Since both the precision of calibration parameters and the computational cost depend on $m$, one can tune the value of $m$ to achieve an optimal balance between precision and computational cost. 

\section{Fast Algorithm to Simulate Visibilities Using Global Sky Model}\label{appfastgsm}

For both traditional self-calibration and the absolute calibration described in this paper, it is crucial to have accurate predictions for the visibilities. This requires simulation of both the contributions of bright point sources and diffuse emission, which can be added together due to the linearity of visibilities. While it is computationally easy to compute the contributions of point sources of known flux, it is much harder to compute visibilities from diffuse emission such as that modeled by the global sky model (GSM,  \citealt{GSM}). Dominated by Galactic synchrotron radiation, this diffuse emission is especially important for the low frequencies and angular resolutions typical of current 21~cm experiments. 

In general, we want to compute visibilities
\begin{align} 
y_{u}(f,t)&=\int s(\hat{\boldsymbol{k}}, f, t) B(\hat{\boldsymbol{k}},f)e^{i \boldsymbol{k}\cdot\boldsymbol{d}_{u}} d\Omega_k, \label{eq:gsmvisibility}
\end{align}
where $s(\hat{\boldsymbol{k}},f, t)$ is the magnitude squared of the global sky model at time $t$ in horizontal coordinates, and $B(\hat{\boldsymbol{k}},f)$ the magnitude squared of the primary beam at a given frequency. Performing the integral by summing over all $n_\text{pix}$ pixels in the GSM takes $\mathcal{O}(n_\text{pix} n_b n_f n_t)$ computational steps, where $n_b$ is the number of unique baselines one simulates, $n_f$ is the number of frequency bins, and $n_t$ is the number of visibilities one simulates for one sidereal day. 

The faster algorithm we describe here takes only $\mathcal{O}(n_\text{pix} n_b n_f)$ steps, by taking advantage of the smoothness of the primary beam as well as the diurnal periodicity in Earth's rotation. It applies only to drift-scanning instruments, so $B(\boldsymbol{k},f,t) = B(\boldsymbol{k},f)$ in horizontal coordinates, and is similar in spirit to the ideas proposed by  \citet{shaw}. 

The key idea is to decompose Equation \ref{eq:gsmvisibility} as follows:
\begin{equation} 
y_{u}(f,t)=
\sum_{\ell,m} a^f_{\ell m}\mathcal{B}_{\ell m}^{uf} e^{i m\phi(t)},
\end{equation}
where each $a_{\ell m}^f$ is a spherical harmonic component of the GSM at a given frequency, and each $\mathcal{B}_{\ell m}^{u f}$ is a spherical harmonic component of $B(\hat{\boldsymbol{k}},f)e^{i \boldsymbol{k}\cdot\boldsymbol{d}_{u}}$, both in equatorial coordinates. In this appendix, we describe precisely how to perform this decomposition and why it decreases the computational cost of calculating visibilities from the GSM.

\subsection{Spherical Harmonic Transform of the GSM}
The GSM of  \citet{GSM} is composed of three HEALPIX maps of size $n_\text{side}$ describing different frequency-independent sky principal components $s^c(\hat{\boldsymbol{k}})$ and the relative weights of each component $w^c(f)$ that encode the frequency dependence. We can decompose the spatial dependence into spherical harmonics,
\begin{align}
a_{\ell m}^c=\int Y^*_{\ell m}(\hat{\boldsymbol{k}})s^c(\hat{\boldsymbol{k}}) d\Omega_k
\end{align}
in $\mathcal{O}(n_\text{pix}^\frac{3}{2})$ steps, due to the advantage of HEALPIX format \citep{healpix}. The frequency dependence of the spherical harmonic coefficients of the sky is given by 
\begin{align}
a^f_{\ell m}=\sum_c a_{\ell m}^c w^c(f),
\end{align}
and the total complexity of computing the coefficients $a^f_{\ell m}$  is $\mathcal{O}(n_\text{pix}^\frac{3}{2}) + \mathcal{O}(n_f)$.


\subsection{Spherical Harmonic Transform of the Beam and Phase Factors}

Next, we would like to compute the spherical harmonics components of $B(\hat{\boldsymbol{k}},f)e^{i \boldsymbol{k}\cdot\boldsymbol{d}_{u}}$:
\begin{align}
\mathcal{B}_{\ell m}^{u f}=\int Y^*_{\ell m}(\hat{\boldsymbol{k}})B(\hat{\boldsymbol{k}},f)e^{i \boldsymbol{k}\cdot\boldsymbol{d}_{u}}d\Omega_k.
\end{align}
Substituting the spherical harmonic decompositions of  $B(\hat{\boldsymbol{k}},f)$ and $e^{i \boldsymbol{k}\cdot\boldsymbol{d}_{u}}$ gives
\begin{align}
\mathcal{B}_{\ell m}^{u f}=&\int Y^*_{\ell m}(\hat{\boldsymbol{k}}) \sum_{\ell' m'}B_{\ell',m'}^{f}Y_{\ell' m'}(\hat{\boldsymbol{k}})\nonumber\\
&\times\sum_{\ell'' m''}4\pi i^{\ell''}j_{\ell''}\left(\dfrac{2\pi f}{c} d_u\right)Y^*_{\ell'' m''}(\hat{\boldsymbol{d}}_u) Y_{\ell'' m''}(\hat{\boldsymbol{k}}) d\Omega_k\nonumber\\
=& \sum_{\ell' m'}  \sum_{\ell'' m''}4\pi i^{\ell''}j_{\ell''}\left(\dfrac{2\pi f}{c} d_u\right)B_{\ell' m'}^{f}Y^*_{\ell'' m''}(\hat{\boldsymbol{d}}_u)\nonumber\\
&\times\int Y^*_{\ell m}(\hat{\boldsymbol{k}})Y_{\ell' m'}(\hat{\boldsymbol{k}}) Y_{\ell'' m''}(\hat{\boldsymbol{k}}) d\Omega_k\nonumber\\
=& \sum_{\ell' m'}  \sum_{\ell'' m''}4\pi i^{\ell''}j_{\ell''}\left(\dfrac{2\pi f}{c} d_u\right)B_{\ell' m'}^{f}Y^*_{\ell'' m''}(\hat{\boldsymbol{d}}_u)\nonumber\\
&\times\sqrt{\frac{\left(2\ell+1\right) \left(2\ell'+1\right) \left(2\ell''+1\right)}{4 \pi }} \nonumber\\
&\times(-1)^m
\left(
\begin{array}{ccc}
\ell & \ell' & \ell'' \\
 0 & 0 & 0 \\
\end{array}
\right) \left(
\begin{array}{ccc}
 \ell & \ell' &\ell'' \\
 -m & m' & m'' \\
\end{array}
\right),
\end{align}
where $j_\ell(x)$ is the spherical Bessel function, $\ell'm'$ represent quantum numbers when expanding the primary beam, $\ell''m''$ represent quantum numbers when expanding $e^{i \boldsymbol{k}\cdot\boldsymbol{d}_{u}}$, and the Wigner-3j symbols are results of integrating the product of three spherical harmonics. Because the Wigner-3$j$ symbols vanish unless $\ell-\ell'\leq\ell''\leq\ell+\ell'$ and $-m+m'+m''=0$, the above sum simplifies to
\begin{align}
\mathcal{B}_{\ell m}^{u f}&= \sum_{\ell' m'}  \sum_{\ell''=\ell-\ell'}^{\ell+\ell'}4\pi i^{\ell''}j_{\ell''}\left(\dfrac{2\pi f}{c} d_u\right)B_{\ell' m'}^{f}Y^*_{\ell'' m''}(\hat{\boldsymbol{d}}_u)\nonumber\\
&\times
\sqrt{\frac{\left(2\ell+1\right) \left(2\ell'+1\right) \left(2
 \ell''+1\right)}{4 \pi }} \nonumber\\
&\times(-1)^m
\left(
\begin{array}{ccc}
 \ell & \ell' & \ell'' \\
 0 & 0 & 0 \\
\end{array}
\right) \left(
\begin{array}{ccc}
 \ell & \ell' & \ell'' \\
 -m & m' & m''\\
\end{array}
\right),
\end{align}
where $m''=m-m'$.
Note that $\ell'$, $m'$ and $\ell''$ in this sum are all limited to the range of $\ell$-values where the spherical harmonics components for the primary beam are non-zero, so the complexity for this triple sum is $n_{B\text{pix}}^\frac{3}{2}$, where $n_{B\text{pix}}$ is the number of non-zero spherical harmonics components for the primary beam. Since the cost for each $\mathcal{B}_{\ell m}^{u f}$ is $n_{B\text{pix}}^\frac{3}{2}$, and there are $n_b n_f n_\text{pix}$ of them, the computational complexity of calculating all $\mathcal{B}_{\ell m}^{u f}$-coefficients scales like $\mathcal{O}(n_b n_f n_\text{pix} n_{B\text{pix}}^\frac{3}{2})$.


\subsection{Computing Visibilities}

By performing a coordinate transformation on Equation \ref{eq:gsmvisibility} from horizontal coordinates 
(corresponding to the local Horizon at the observing site)
to equatorial coordinates, the time dependence of $s(\hat{\boldsymbol{k}})$ is transferred to $B(\hat{\boldsymbol{k}})$ and $\boldsymbol{d}_u$. We can now calculate $y_u(f,t)$ by computing
\begin{align} 
y_u(f,t)&=\int s(\hat{\boldsymbol{k}}) B^{f t}(\hat{\boldsymbol{k}})e^{i \boldsymbol{k}\cdot\boldsymbol{d}_u(t)} d\Omega_k\nonumber\\
&=\sum_{\ell m} a^*_{\ell m}\mathcal{B}_{\ell m}^{u f t}.
\end{align}
Since the time dependence of $\mathcal{B}_{\ell m}^{u f t}$ is a constant rotation along the azimuthal direction, we can write the above as 
\begin{equation} 
y_u(f,t)=\sum_{\ell m} a^*_{\ell m}\mathcal{B}_{\ell m}^{u f} e^{i m\phi(t)}=\sum_{m}c_m^{uf} e^{i m\phi(t)}, \label{eq:fftvisibility}
\end{equation}
where we have defined 
\begin{equation} 
c_m^{u f} \equiv \sum_{\ell} a^*_{\ell m}\mathcal{B}_{\ell m}^{u f},
\end{equation}
which can be computed in $\mathcal{O}(n_b n_f n_\text{pix})$ steps. Given $c_m^{u f}$, it is clear that we can evaluate Equation \ref{eq:fftvisibility} using a fast Fourier Transform (FFT), whose cost is
\begin{align} 
\mathcal{O}(n_b n_f n_t \text{log}(n_t)).
\end{align}
 Note that this FFT in Equation \ref{eq:fftvisibility} has no $n_\text{pix}$ dependence, because we always need to zero-pad $c_m$ to length $n_t$ before the FFT.
%
In summary, the total complexity of all of the above steps is 
\begin{align} 
&\mathcal{O}\left(n_\text{pix}^\frac{3}{2}\right) + \mathcal{O}(n_f)+\mathcal{O}\left(n_b n_f n_\text{pix} n_{B\text{pix}}^\frac{3}{2} \right) \nonumber\\ 
&+\mathcal{O}\left(n_b n_f n_\text{pix}) +\mathcal{O}(n_b n_f n_t \text{lg}(n_t)\right)\nonumber\\
\approx& \mathcal{O}\left(n_b n_f n_\text{pix} n_{B\text{pix}}^\frac{3}{2} \right).
\end{align}
This does not scale with $n_t$, unlike the naive integration's $\mathcal{O}(n_b n_f n_\text{pix} n_t)$. Thus with a spatially smooth beam whose $ n_{B\text{pix}} \ll n_t^\frac{2}{3}$, the algorithm described here is  much faster than the naive numerical integration approach described at the beginning of this Appendix.

\end{document}